\documentclass[amsmath,amssymb,aps,prd,11pt,tightenlines,superscriptaddress, 
nofootinbib,preprintnumbers,notitlepage]{revtex4-1}
\usepackage{graphicx}
\usepackage{dcolumn}
\usepackage{bm}
\usepackage{amssymb}
\usepackage{amsmath}
\usepackage{mathrsfs}
\usepackage{graphicx,color} 
\usepackage[english]{babel}     
\usepackage{graphicx,wrapfig,lipsum}      
\usepackage{multimedia}    
\usepackage[mathscr]{eucal}  
\usepackage{bm}              
\usepackage{amssymb}           
\usepackage{amsmath}      
\usepackage{mathrsfs}  
\usepackage[mathscr]{eucal}      
\usepackage[normalem]{ulem}
\usepackage{tikz}
\usepackage{cleveref}
\makeatletter
\newcommand*{\encircled}[1]{\relax\ifmmode\mathpalette\@encircled@math{#1}\else\@encircled{#1}\fi}
\newcommand*{\@encircled@math}[2]{\@encircled{$\m@th#1#2$}}
\newcommand*{\@encircled}[1]{%
  \tikz[baseline,anchor=base]{\node[draw,circle,outer sep=0pt,inner sep=.2ex] {#1};}}
\makeatother

\begin{document}
\title{Prospecting effective Yang-Mills-Higgs models for the asymptotic confining flux tube}  
\affiliation{
Instituto de F\'\i sica, Universidade Federal Fluminense, 24210-346 Niter\'oi - RJ, Brasil.} 
\affiliation{
Instituto de F\'isica Te\'orica, Universidade Estadual Paulista, 01140-070 S\~ao Paulo - SP, Brasil.} 
\author{David R. Junior}
\affiliation{
Instituto de F\'isica Te\'orica, Universidade Estadual Paulista, 01140-070 S\~ao Paulo - SP, Brasil.} 
\affiliation{
Instituto de F\'\i sica, Universidade Federal Fluminense, 24210-346 Niter\'oi - RJ, Brasil.} 
\author{Luis E. Oxman}
\affiliation{
Instituto de F\'\i sica, Universidade Federal Fluminense, 24210-346 Niter\'oi - RJ, Brasil.} 
\author{Gustavo M. Sim\~oes}  
\affiliation{
Instituto de F\'\i sica, Universidade Federal Fluminense, 24210-346 Niter\'oi - RJ, Brasil.}

\date{\today} 

\begin{abstract}

In this work, we analyze a large class of effective Yang-Mills-Higgs  models constructed in terms of adjoint scalars.  In particular, we reproduce  asymptotic properties of the confining string, suggested by lattice simulations of $SU(N)$ pure Yang-Mills theory, in models that are stable in the whole range of Higgs-field mass parameters. These properties include $N$-ality, Abelian-like flux-tube profiles, independence of the 
profiles with the $N$-ality of the quark representation, and Casimir scaling. We find that although these models are formulated in terms of many fields and possible Higgs potentials, a collective behavior can be established in a large region of parameter space, where the desired asymptotic behavior is realized. 

\end{abstract}

\maketitle
\section{Introduction}
\label{Intro}
The dual superconductivity scenario to describe confinement in pure Yang-Mills (YM) theory has been a subject of intense research for several decades \cite{superconduct-1,superconduct-2,superconduct-3,superconduct-4,superconduct-5,superconduct-6}. According to this mechanism, the Yang-Mills vacuum behaves as a condensate of chromomagnetic objects that gives rise 
to a confining flux tube between quark probes. This idea has been explored extensively using lattice simulations. For example, along the transverse direction to the flux tube, the profile for the longitudinal component of the chromoelectric field has been fitted with the solitonic Abelian 
Nielsen-Olesen vortex \cite{abelian-like}. The underlying objects that could condense in four-dimensional spacetime have also been studied \cite{cv-1,cv-2,cv-3,cv-4,cv-5,cv-6,cv-7,cv-8,cv-9,cv-10,cv-11}. Ensembles formed by monopoles that propagate along worldlines and thin center-vortices, which are gauge field configurations characterized by loops that propagate along worldsurfaces, have been identified in the YM vacuum  \cite{cv-6}. In particular, the $N$-ality property observed in large Wilson loops was reproduced when the average over Monte Carlo configurations is replaced by one over simpler thin center-vortex configurations,  extracted from the complete link variables, which happen to percolate in the continuum limit. Then, one important question is how to conciliate the Abelian-like behavior of the flux tube and $N$-ality. These features can be accommodated in effective Yang-Mills-Higgs (YMH) field models with $SU(N) \to Z(N)$ spontaneous symmetry breaking (SSB) \cite{effmodels-1,effmodels-2,effmodels-3,effmodels-4,effmodels-5,effmodels-6,effmodels-7,effmodels-8}, with or without $SU(N)$ flavor symmetry. Moreover, $SU(N)$ gauge field models constructed with adjoint Higgs fields effectively describe the asymptotic behavior of the different condensates observed in the lattice  \cite{4densemble} (see also \cite{ymvacuum}).  In these models, the effective $SU(N)$ gauge field $\Lambda_\mu$ represents the Goldstone modes for the percolating thin center vortices, with the natural $N$-matching rule among center-vortex worldsurfaces. The adjoint Higgs fields, minimally coupled to $\Lambda_\mu$,  effectively describe monopole worldlines attached to worldsurfaces, thus including the nonoriented (in the Lie algebra) center-vortex component. The Higgs potential contains 
a mass term $m^2 (\psi_I, \psi_I)$ and the natural matching rules among monopole worldlines. On this direction, an $SU(N)$
color and flavor symmetric model based on adjoint fields $\psi_I \in \mathfrak{su}(N)$ with $N^2-1$ flavors, $I= 1, \dots, N^2-1$, was analyzed in Refs. \cite{valencegluons,model-1,model-2}. Within this framework, when $m^2=0$, the flux tube between external probes coincides with the Abelian Nielsen-Olesen (finite) vortex. In addition, at asymptotic distances, as the group representation of the external probes is varied, the string tension satisfies a Casimir scaling law.

From a phenomenological point of view, studies in the lattice cast some doubts about whether field models with Nielsen-Olesen profiles 
are suitable to describe the confining string. The analysis of the energy-momentum tensor showed deviations from the Abelian counterparts at intermediate distances \cite{emt-ym}. A possible way out could be the consideration of non-Abelian models away from the Abelianization point. However, it 
could also happen that the intermediate confining regime lies outside the domain of applicability of the effective field model. Being originated from thin objects, it could only be used at asymptotic distances.
In this case, there is still an issue with the model studied in Ref. \cite{model-1}: when moving from the Abelianization point at $m^2=0$, there is a neighboring region ($m^2<0$) where the model becomes unstable. In fact, it would be interesting if stability could be realized with negative $m^2$, which is naturally obtained in ensembles where monopole worldlines have negative tension (monopole proliferation) and positive stiffness  \cite{mono-1,mono-2}.  On this direction, this state could be stabilized by additional quartic terms not considered in the original formulation. Indeed, the model described in Ref. \cite{model-1} does not contain all possible terms compatible with color and flavor symmetry. In this work, we present a thorough investigation of the most general flavor-symmetric $SU(N)$ model with $N^2-1$ adjoint Higgs flavors, studying the possibility of coexistence of asymptotic $N$-ality, Abelian-like profiles, Casimir scaling, and stable regions in parameter space. These are important properties compatible with present lattice simulations of pure YM theory. The observed independence of the flux-tube cross-section with respect to the $N$-ality of the quark representation \cite{stability} will also be discussed.

 \section{$SU(N)$ effective models with adjoint scalars}

Lattice simulations showed that oriented (in the Lie algebra) and nonoriented center vortices are relevant  variables to describe the various aspects of the infrared  behavior of $SU(N)$ Yang-Mills theory  (for a recent review,  see Ref. \cite{review}). In Ref. \cite{4densemble}, it was shown that the effective description of a 4d mixed ensemble  of percolating oriented and nonoriented center-vortex worldsurfaces is given by a Yang-Mills-Higgs model with a Higgs content based on a set of adjoint scalars. The ensemble was initially defined on the lattice, with the oriented component generated by means of a Wilson action for dual $SU(N)$ link variables $V_\mu$,
\begin{align}
    S_{\rm eff}=\zeta \,{\rm Re}\,{\rm tr}\left[I-V_\mu(x)V_\nu(x+\hat{\mu})V_\mu^\dagger(x+\hat{\nu})V_\nu^\dagger(x)e^{-i\alpha_{\mu\nu}(x)}\right]\;,
    \label{dwilson}
\end{align}
with frustration $e^{-i\alpha_{\mu\nu}(x)}$. 
 The latter is nontrivial on plaquettes that intersect a surface $S(C)$ whose border is the Wilson loop $C$ and is trivial otherwise. The nontrivial value is given by
 \[
   e^{-i\alpha_{\mu\nu}} = e^{-i2\pi \beta_{\rm e}|_q\ T_q}\;,
   \]
 which is the center-element generated when an elementary center-vortex worldsurface links the Wilson loop for quarks in representation $D(\cdot)$. Here, $T_q$, $q=1, \dots, N-1$, are the Cartan generators and the $(N-1)$-tuple $\beta_{\rm e}$ is a magnetic weight of the quark representation.  Indeed, when 
an expansion in powers of $\zeta$ is performed, because of the properties of the Haar measure, a nonvanishing contribution arises from links on plaquettes that form closed surfaces. Because of the frustration, these contributions are accompanied by a center element when the vortex surfaces link the Wilson loop. Moreover, as the group is $SU(N)$, plaquette configurations where $N$ vortex surfaces are attached to a curve also contribute. These are the fingerprints of center-vortices. In that reference, nonoriented configurations where pairs of worldsurfaces are attached to closed monopole worldlines, as well as the natural three and four-line rules to match open monopole worldlines at a common endpoint, were also introduced. This was done by  including a diluted gas of adjoint loop hololomies together with  structures formed by holonomies along open lines with various natural matching rules among the possible adjoint charges of monopoles (the roots of $\mathfrak{su}(N)$).  In the naif continuum limit, Eq. \eqref{dwilson} gives rise to 
 a dual $SU(N)$ gauge field $\Lambda_\mu$ governed by a Yang-Mills action $ (F_{\mu\nu}(\Lambda)-J_{\mu\nu})^2$ with frustration $J_{\mu \nu}$ proportional to $\beta_{\rm e}$ and localized on $S(C)$. In the gas of adjoint monopole loops and structures formed by three and four matched lines, each holonomy gets replaced by the continuum version
 \begin{align}
\Gamma_\gamma[\Lambda]={\rm Ad}\left(\exp\left[\int_\gamma \Lambda_\mu dx^\mu\right]\right)\;, 
\end{align}
where $\gamma$ is a monopole worldline. 
These lines were equipped with tension $\mu$ and positive stiffness $1/\kappa$. Using polymer techniques, it was shown that the gas of adjoint loop holonomies 
lead to an effective description in terms of adjoint scalars $\psi_I$ ($I$ is a flavor label) with squared mass $m^2 \propto \mu \kappa$, which are minimally coupled to the dual gauge field $\Lambda_\mu$. In addition, the three and four-line structures correspond to cubic and quartic interaction terms for these scalars.  In this setting, the emergence of $N(N-1)$ flavors is immediate, as this is the number of adjoint monopole charges (roots) in the Lie algebra $\mathfrak{su}(N)$. The consideration of additional adjoint scalars $\psi_q$, $q=1, \dots , N-1$, could also be useful to implement other matching rules as well as a color and flavor symmetry in an effective model with $N^2-1$ flavors. As lattice simulations have not yet constrained the possible correlations between the  monopoles, we will study a large class of models with or without color and flavor symmetry. In the former case, all cubic and quartic terms compatible with color and flavor symmetry shall be considered. Among the possibilities,  the model studied   in Refs. \cite{valencegluons,model-1,model-2} is an interesting particular case that is characterized by just one cubic and one quartic term. At $m^2=0$ and asymptotic distances, it displays
$N$-ality, exact Casimir scaling, and Abelian-like profiles independent of the quark representation, properties which are compatible with those observed in the lattice for the confining flux tube. Nonetheless, it could well happen that future simulations regarding the properties of nonoriented center vortices favor a mass parameter $m^2<0$. In this case, the model studied in Refs. \cite{valencegluons,model-1,model-2}
could not be applicable because, in that region, the potential energy is not bounded from below. Note that a negative squared mass is a common feature in monopole-only lattice scenarios based on the Abelian projection, as it effectively describes the observed monopole proliferation (monopole condensate) \cite{mono-1,mono-2,mono-3}.   For these reasons, we will look for the possibility of  stable generalized models, which are consistent when $m^2 < 0$ and that keep the above mentioned interesting asymptotic properties of the confining flux tubes.

\section{General $SU(N)$ model with $N^2-1$ flavors}
\label{GeneralSUNmodel}

In this section, we shall initially review the effective $SU(N)$ YMH model with $N^2-1$ adjoint scalar fields proposed in Ref. \cite{conf-qg}. Its action reads
\begin{align}
 &S=\int d^4x \left( \frac{1}{4} \langle F_{\mu\nu},F_{\mu\nu}\rangle + \frac{1}{2}\langle D_\mu \psi_I, D_\mu \psi_I \rangle + V_{\rm H}(\psi) \right)  \makebox[.5in]{,} \psi_I \in \mathfrak{su}(N) \;,\notag\\    
&F_{\mu\nu} =\frac{i}{g} \left[D_\mu,D_\nu\right]\makebox[.5in]{,}  D_\mu =\partial_\mu +g \Lambda_\mu\wedge \;, \notag\\ 
&V_{\rm H} (\psi) = c+\frac{\mu^2}{2}\langle \psi_A,\psi_A\rangle 
+ \frac{\kappa}{3}f_{ABC}\langle \psi_A \wedge \psi_B, \psi_C\rangle 
+ \frac{\lambda}{4}\langle \psi_A\wedge \psi_B,\psi_A\wedge \psi_B\rangle\;.
\label{higgs} 
\end{align} 
Here, we used the notation $X\wedge Y=-i[X,Y]$, while the brackets denote the Killing product between two Lie Algebra elements, defined by
\begin{align}
    \langle X,Y\rangle = {\rm Tr}  \left( {\rm Ad}(X) {\rm Ad}(Y)\right)\makebox[.5in]{,} X,Y\in\mathfrak{su}(N) \;,
\end{align}
where $ {\rm Ad}(X) $ refers to the adjoint representation.   In our conventions, the basis ${T_A}$ satisfies $\langle T_A, T_B\rangle=\delta_{AB}$. This model is invariant under color transformations
 \begin{align}
     \psi_I\to U\psi_I  U^{-1} \makebox[.5in]{,} U\in SU(N)
     \label{colortransf}
 \end{align}
 and under flavor transformations
 \begin{align}
    \psi_I\to {\rm Ad}(U)_{IJ}\psi_J\;.\label{flavortransf}
 \end{align}
For an appropriate choice of the parameters, an $SU(N) \to Z(N)$ spontaneous symmetry breaking (SSB) pattern is triggered. Then, as the first homotopy group of the vacuum manifold $\mathcal{M}=SU(N)/Z(N)$ is $Z(N)$, the topologically stable vortex solutions display $N$-ality. At $m^2=0$, they are Abelian-like and have an exact Casimir law at asymptotic distances for the $k$-Antisymmetric representations \cite{OxmanSimoes2019}. Furthermore, at $\lambda=g^2$,  this scaling law was shown to be stable as the energy of the $k$-Antisymmetric irrep is the smallest among the irreps with $N$-ality $k$ \cite{JuniorOxmanSimoes2020}. These properties are compatible with those of the confining string observed in lattice simulations \cite{stability,casimir-4d}.  Note that the $SU(N)\to Z(N)$ SSB becomes unstable for $m^2<0$, as the energy of aligned configurations ($\psi_A= \psi_B\,$ for all $A$, $B$) is arbitrarily negative for large $\langle \psi_A , \psi_A \rangle$. This happens because both the cubic and the quartic terms are zero in this case. Thus, we are led to look for new models with additional relevant terms to stabilize the desired phase. Initially, it is interesting to consider a potential that depends on $\psi_A$ through the real variable 
\begin{align}
     \phi(x,g)=\langle \psi_A, g T_A g^{-1}\rangle \makebox[.5in]{,} g\,\in SU(N) 
 \end{align}
 as follows
\begin{eqnarray}
\label{grouppotential}
   &V(\psi)=c_0+\int d \mu(g)\,\left( \frac{a}{2}\phi^2+\frac{b}{3}\phi^3+\frac{c}{4}\phi^4 \right) 
    \;.  \label{groupm}
\end{eqnarray}
This potential can be shown to be invariant under both color and flavor transformations (cf. eqs. \eqref{colortransf}, \eqref{flavortransf}) after using the invariance of the Haar measure $\int d\mu(g)=\int d\mu(gU)$, $ U \in SU(N)$. Here, it is also clear that the quartic term is non-negative and the potential is bounded from below. In particular, that term would vanish only if $\phi(g)$ itself vanished for every $U$, in which case the quadratic term would also be zero. Therefore, the above-mentioned stability issue does not exist in this case. 
Indeed,  when compared with eq. \eqref{higgs}, this model generates additional terms, which happen to be all possible terms that are compatible with color and flavor symmetry up to quartic order, given in a specific combination. In what follows, we will show this statement while, in the next section, we will study the most general model with this symmetry, obtained by assigning arbitrary coefficients to all generated terms.

\begin{figure}
    \centering
    \includegraphics[width=0.8\textwidth]{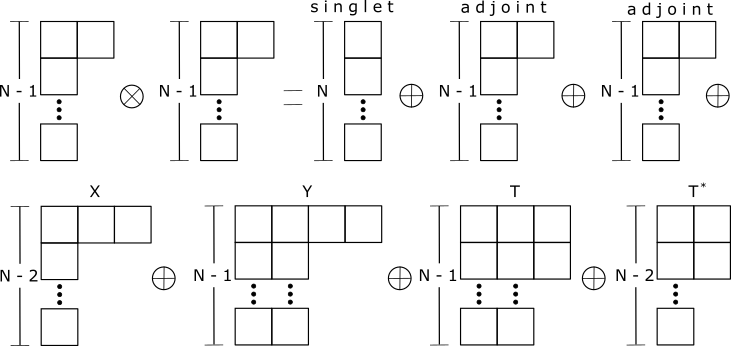}
    \caption{Young Tableaux of all irreps. contained in the tensor product of two $Ad(SU(N))$ representations.}
    \label{AdxAd}
\end{figure}

Since $\psi(g)=R_{AA'}(g)\psi_{AA'}$, where $R(g)$ stands for the adjoint representation matrix of $g$, the integrand contains two, three and four tensor products of adjoint representations of $SU(N)$. Only the singlet part of these tensor products yields a nonvanishing contribution for the integral. According to Refs. \cite{Chivukula,Keppeler}, the tensor product of two adjoint representations $R(g)_{ab}R(g)_{cd}$ can be decomposed into 7 different irreps\footnote{This is for $N>3$. For $N=2$, only the representations $S$, $F$ and $X$ exist and are associated with spins $0$, $1$, and $2$ respectively. For $N=3$, the representation $Y$ does not exist. Notice that this does not mean that the corresponding projectors vanish if one naively sets $N=2$ or $N=3$. Simply, they should not be considered in these particular cases.} (the Young tableaux are shown in Fig. \ref{AdxAd}). The associated Hermitian projectors are
\begin{subequations}
\label{Projectors}
\begin{align}
   P_S\vert^{AB}_{CD} =&\frac{1}{N^2-1}\delta_{AB}\delta_{CD}\;,\\
    P_F\vert^{AB}_{CD} =&f_{ABE}f_{CDE}\;,\\
    P_D\vert^{AB}_{CD} =&\frac{N^2}{N^2-4}d_{ABE}d_{CDE}\;,\\
    P_X\vert^{AB}_{CD} =&\frac{1}{4}\delta_{AC}\delta_{BD}+\frac{N+2}{4N}\delta_{AD}\delta_{BC}-\frac{1}{2N(N+1)}\delta_{AB}\delta_{CD}\notag\\
    &-\frac{N}{4}f_{ADE}f_{BCE}+\frac{N}{4}d_{ADE}d_{BCE}-\frac{N}{2(N+2)}d_{ABE}d_{CDE}\;,\\
    P_Y\vert^{AB}_{CD} =&\frac{N-2}{4N}(\delta_{AC}\delta_{BD}+\delta_{AD}\delta_{BC})+\frac{N-2}{2N(N-1)}\delta_{AB}\delta_{CD}\notag\\
    &-\frac{N}{4}(d_{ACE}d_{BDE}+d_{ADE}d_{BCE})+\frac{N(N-4)}{4(N-2)}d_{ABE}d_{CDE}\;,\\
    P_T\vert^{AB}_{CD} =&\frac{1}{4}(\delta_{AC}\delta_{BD}-\delta_{AD}\delta_{BC})-\frac{1}{2}f_{ABE}f_{CDE}+i\frac{N}{4}(f_{ADE}d_{BCE}+d_{ADE}f_{BCE})\;,\\
    P_{\overline{T}}\vert^{AB}_{CD}=&P_T^*\vert^{AB}_{CD}\;,
\end{align}
\end{subequations}
where the symmetryc and antisymmetryc structure constants $d_{ABC}$ and $f_{ABC}$ are defined as
\begin{subequations}
    \begin{align}
        T_A\wedge T_B &= f_{ABC}T_C\;,\\
       \{T_A,T_B\}&=\frac{\delta_{AB}}{N^2}\mathbb{I} +d_{ABC}T_C\;.
    \end{align}
\end{subequations}
These projectors allow us to obtain explicit expressions for the different integrals by decreasing the number of matrices in the integrand up to a point where the  orthogonality relations \cite{Hamermesh},
\begin{equation}
\label{ortho-irrep}
    \int d\mu(g) D^{(i)}(g)\vert_{\xi\xi'}D^{(j)}(g^{-1})\vert_{\zeta'\zeta} = \frac{\delta_{ij}\delta_{\xi\zeta}\delta_{\xi'\zeta'}}{d_i}\;,
    \end{equation}
can be used. The indices $i,j$ label irreducible representations,  while $d_i$ is the dimension of $D^{(i)}$. Since these representations are unitary, their components satisfy $D(g^{-1})_{B'B}=D(g)^\dagger_{B'B}=D^*(g)_{BB'}$. 
For the quadratic term, we can directly use Eq, \eqref{ortho-irrep} to obtain
\begin{equation}
    \int d\mu(g)\, \phi^2 = \langle\psi_A,T_{A'}\rangle\langle\psi_B,T_{B'}\rangle \int d\mu(g) R(g)_{AA'}R(g)_{BB'} = \frac{\langle \psi_A,\psi_A\rangle}{N^2-1}  \;.
    \label{quad}
\end{equation}
To evaluate the cubic term, the completeness property
\begin{eqnarray}
  \delta_{A A'}\delta_{B B'}  = P_S\vert_{A'B'}^{A B}+P_F\vert_{A'B'}^{A B}+\dots  
\end{eqnarray}
leads to
\begin{align}
    &I_3\vert^{ABC}_{A'B'C'} = \int d\mu(g) R(g)_{AA'}R(g)_{BB'}R(g)_{CC'}\notag\\
    &=\int d\mu(g) R(g)_{AA''}R(g)_{BB''}\left(P_S\vert^{A''B''}_{A'B'}+P_F\vert^{A''B''}_{A'B'}+P_D\vert^{A''B''}_{A'B'}+...\right)R(g)_{CC'}\label{last}\\
    &=f_{ABC}f_{A'B'C'}+\frac{N^2}{N^2-4}d_{ABC}d_{A'B'C'}\;.
\end{align}
Here, we used that $P_i$ selects the subspace carrying the irreducible representation $i$ so that, in each term, the product $R_{AA'}R_{BB'}P_{(i)}\vert^{A''B''}_{A'B'}$ can be thought of as components of a single $D^{(i)}$, which allows us to use the orthogonality relations to evaluate the group integral. As the last factor in eq. \eqref{last} is in the adjoint, the only contribution is originated from the two independent subspaces that carry an adjoint representation in ${\rm Ad}\otimes {\rm Ad} $. 
In this manner, we get
\begin{eqnarray}
\label{IntegratedCubic}
     \int d\mu(g)\, \phi^3 =\frac{f_{ABC}\langle\psi_A\wedge\psi_B,\psi_C\rangle}{N^2-1}+\frac{N^2d_{ABC}\langle \psi_A\vee\psi_B,\psi_C\rangle}{(N^2-1)(N^2-4)}\;.\notag\\ \label{cubicgroup}
\end{eqnarray}
where we defined
\begin{align}
    X\vee Y = \{X,Y\}-\frac{\langle X,Y\rangle}{N^2}\;.
\end{align}
Similarly, We can proceed with the quartic term when computing
\begin{align}
    &I_4\vert^{ABCD}_{A'B'C'D'} = \int d\mu(g) R(g)_{AA'}R(g)_{BB'}R(g)_{CC'}R(g)_{DD'} \;.
\end{align}
This time we can introduce a pair of completeness relations to reduce the components of $R(g)$ in terms of components of $D^{(i)}$ and then use the orthogonality relation in eq. \eqref{ortho-irrep}. Most of the contributions will be originated from products reduced by the same projectors. 
Two notable exceptions are the products between the representations $T$ and $\bar{T}$ and the two different adjoints, one associated with $f_{ABC}$ and the other with $d_{ABC}$. The result for the quartic term is
\begin{subequations}
\begin{align}
\label{IntegratedQuartic}
    &\int d\mu(g)\, \phi^4 =\langle\psi_A,T_{A'}\rangle\langle\psi_B,T_{B'}\rangle\langle\psi_C,T_{C'}\rangle\langle\psi_D,T_{D'}\rangle I_4\vert^{ABCD}_{A'B'C'D'}\;,\\
    &I_4\vert^{ABCD}_{A'B'C'D'}=\frac{N^2f_{ABE}d_{CDE}f_{A'B'E'}d_{C'D'E'}+N^2d_{ABE}f_{CDE}d_{A'B'E'}f_{C'D'E'}}{(N^2-1)(N^2-4)}+\notag\\
    &+\frac{4P_T\vert^{AB}_{CD}P_{\overline{T}} \vert^{A'B'}_{C'D'}+4P_{\overline{T}}\vert^{AB}_{CD}P_T \vert^{A'B'}_{C'D'}}{(N^2-1)(N^2-4)}+\sum\limits_{i=S,F,D,X,Y}\frac{1}{d_i}P_i\vert^{AB}_{CD}P_i\vert^{A'B'}_{C'D'}\;.
\end{align}
\end{subequations}

Now, let us see that the terms generated by the model in eq. \eqref{groupm} are all possible terms invariant under the desired symmetry. It is clear that the quadratic term on the right-hand side of eq. \eqref{quad} is the only possibility. As for the cubic term, the most general combination is given by
\begin{align}
    C^{ABC}_{A'B'C'}\psi_{AA'}\psi_{BB'}\psi_{CC'}\;,
\end{align}
with $\psi_{AA'}=\langle\psi_A,T_{A'}\rangle$, i.e. primed indices refer to color and unprimed refer to flavor. To ensure that color and flavor symmetries
\begin{align}
    \psi_{A A'}\to R_{A' B'}\psi_{A B'} \makebox[.5in]{,}     \psi_{A A'}\to R_{A B}\psi_{B A'}\;,
\end{align}
are independently present, 
$ C^{ABC}_{A'B'C'}$ must be a linear combination of the antisymmetric and symmetric structure constants of $\mathfrak{su}(N)$ in both sets of prime and unprimed indices.  Indeed, these structure constants are the only invariant tensors with three indices, that is, the only singlets in ${\rm Ad}^{\otimes 3} \equiv {\rm Ad}\otimes {\rm Ad} \otimes {\rm Ad}$. Therefore,  the most general cubic term can be parametrized by\begin{align}
    \frac{\kappa_f}{3}f_{ABC}\langle \psi_A\wedge\psi_B,\psi_C\rangle + \frac{\kappa_d}{3} d_{ABC}\langle \psi_A \vee\psi_B,\psi_C\rangle \;,
\end{align}
which corresponds to assigning arbitrary coefficients to the terms obtained in eq. \eqref{cubicgroup}. Regarding the quartic term, the most general possibility is
\begin{eqnarray}
C^{ABCD}_{A'B'C'D'}\psi_{AA'}\psi_{BB'}\psi_{CC'}\psi_{DD'}\;.
\end{eqnarray}
This time, $C^{ABCD}_{A'B'C'D'}$ must be a linear combination of terms of the form 
\begin{eqnarray}
     T_{\rm f}^{ABCD}T_{\rm c}^{A'B'C'D'}\;,
\end{eqnarray}
 with both tensors $T_{\rm f}$ and $T_{\rm c}$ being invariant under the adjoint action of $ SU(N)$.   The space of invariants with four adjoint indices (singlets in ${\rm Ad}^{\otimes 4}$) has basis\footnote{In fact, for $SU(2)$ and $SU(3)$ this basis is overcomplete and should include only $3$ and $8$ elements, respectively. For $SU(2)$, $d_{ABC}=0$ and $f_{ABE}f_{CDE} = \delta_{AC}\delta_{BD}-\delta_{AD}\delta_{BC}$. For $SU(3)$, Burgoyne's identity $3(d_{ABE}d_{CDE}+d_{ACE}d_{BDE}+d_{ADE}d_{BCE}) = \delta_{AB}\delta_{CD}+\delta_{AC}\delta_{BD}+\delta_{AD}\delta_{BC}$ can be used.} $\delta_{AB}\delta_{CD}$, $\delta_{AC}\delta_{BD}$, $\delta_{AD}\delta_{BC}$, $d_{ABE}d_{CDE}$, $d_{ACE}d_{BDE}$, $d_{ADE}d_{BCE}$, $f_{ABE}f_{CDE}$, $f_{ACE}f_{BDE}$, and $f_{ADE}f_{BCE}$ \cite{Chivukula}. In principle, with $9$ invariants, the most general $C^{ABCD}_{A'B'C'D'}$ is written as a linear combination of $81$ interactions. However, at most 9 of these 81 interactions\footnote{the number drops to 3 and 8 for $SU(2)$ and $SU(3)$ respectively.} are non-vanishing and linearly independent. The key equations to eliminate these redundancies are
\begin{eqnarray}
     f_{ABE}f_{CDE}&=&d_{ACE}d_{BDE}-d_{ADE}d_{BCE}+\frac{2}{N^2}(\delta_{AC}\delta_{BD}-\delta_{AD}\delta_{BC})\;,\label{fdIdentity1}\\
     f_{ABE}d_{CDE}&=&d_{ADE}f_{BCE}+d_{ACE}f_{BDE}\;.\label{fdIdentity2}
\end{eqnarray}
For organizational purposes, these 9 interactions can be subdivided into four sets, depending on how the flavor indices of the Higgs fields are matched. In the flavor-singlet set, the flavor indices are contracted with each other:
\begin{subequations}
\begin{align}
    V_1^{(4)}=(\langle\psi_A,\psi_A\rangle)^2\;& \makebox[.5in]{;}   V_2^{(4)}=\langle\psi_A,\psi_B\rangle\langle\psi_A,\psi_B\rangle\;;\\
    V_3^{(4)}=\langle\psi_A\wedge\psi_B , \psi_A\wedge\psi_B\rangle\;& \makebox[.5in]{;}   V_4^{(4)}=\langle\psi_A\vee\psi_A,\psi_B\vee\psi_B\rangle\;.
\end{align}
\end{subequations}
Notice $V_3$ is the interaction that was analyzed in previous papers \cite{conf-qg,OxmanSimoes2019,JuniorOxmanSimoes2020}. In the f-adjoint set, the flavor indices are matched with two copies of the antisymmetric  structure constants $f_{ABC}$:
\begin{subequations}
    \begin{align}
        V_5^{(4)}&=f_{ABE}f_{CDE}\langle \psi_A,\psi_C \rangle\langle\psi_B,\psi_D\rangle\;;\\
    V_6^{(4)}&=f_{ABE}f_{CDE}\langle\psi_A\wedge\psi_B ,\psi_C\wedge\psi_D \rangle\;.\label{V6}
    \end{align}
\end{subequations}
The d-adjoint set is analogous to the f-adjoint one, but with symmetric structure constants $d_{ABC}$ instead:
\begin{subequations}
    \begin{align}
        V_{7}^{(4)}&=d_{ABE}d_{CDE}\langle\psi_A,\psi_B\rangle\langle\psi_C,\psi_D\rangle\;;\\
    V_{8}^{(4)}&=d_{ABE}d_{CDE}\langle\psi_A\vee\psi_B,\psi_C\vee\psi_D\rangle\;,
    \end{align}
\end{subequations}

Finally, there is the mixed adjoint set, where the flavor indices are matched with both $f_{ABC}$ and $d_{ABC}$:
\begin{subequations}
    \begin{align}
            V_{9}^{(4)}&=f_{ABE}d_{CDE}\langle\psi_A\wedge\psi_B,\psi_C\vee\psi_D\rangle\;,\\
    \end{align}
\end{subequations}
These terms are all generated in eq. \eqref{IntegratedQuartic}. 

\section{Ansatz for the vortex solutions}

In this section, we shall study the most general color and flavor symmetric model for a set of $N^2-1$  SU(N) adjoint Higgs fields. The total energy in the presence of static $\Lambda_0=0$ gauge fields is 
\begin{eqnarray}
    E=\int d^4x\,\left(\frac{1}{2}\langle B_i,B_i\rangle +\langle D_\mu(\Lambda)\psi_I,D_\mu(\Lambda)\psi_I\rangle + V_{\rm gen}(\psi)\right)\;.\label{gen_model}
\end{eqnarray}
 The most general potential, as discussed in the previous section, is given by 
\begin{eqnarray}
\label{generalpotential}
     V_{\rm gen}(\psi)=c_0+\frac{1}{2}m^2V^{ (2)}+\frac{1}{3}\kappa_fV^{
(3)-f}+\frac{1}{3}\kappa_dV^{(3)-d}+\frac{1}{4}\sum\limits_{i=1}^9\lambda_i V_i^{(4)}(\psi)\;.\label{genV}   
\end{eqnarray}
Here, we defined
\begin{align}
    &V^{ (2)}=\langle\psi_A,\psi_A\rangle\;,\nonumber\\&V^{(3)-f}=f_{ABC}\langle\psi_A\wedge\psi_B,\psi_C\rangle\;,\nonumber\\& V^{(3)-d}= d_{ABC}\langle\psi_A\vee\psi_B,\psi_C\rangle\;.
\end{align}

 Regarding the stability of the potential in eq. \eqref{generalpotential}, all of the interactions $V^{(4)}_i$ are positive, except for $V^{(4)}_9$, which was verified numerically to have an indefinite sign. However, it is clear that this term can be present in a stable model as it occurs in the one defined in eq. \eqref{grouppotential}. Moreover, the instability problem pointed out in Sec. I when $m^2<0$ can be easily overcome in a large region of parameter space while keeping  Abelian-like behavior and Casimir Scaling. This  will be revisited at the end of Secs. \ref{AbelianizationPoint} and \ref{Modelwitoutpsiq}.

In Ref. \cite{OxmanSimoes2019} an ansatz for a vortex with charge $k$ was proposed for the model defined by eq. \eqref{higgs}. In this subsection, we will show that the same ansatz closes the equations of motion for the general model of eq. \eqref{gen_model}. The ansatz is proposed in the Cartan-Weyl basis of the Lie Algebra, which consists of $N-1$ diagonal generators $T_q,\, q=1,\dots,N-1$, and $\frac{N(N-1)}{2}$ pairs of off-diagonal generator $T_{\alpha}, T_{\bar{\alpha}}$, which are labeled by the positive roots $\alpha$ of $SU(N)$ \footnote{See Appendices A, B for a review of some aspects of the Lie Algebra of SU(N) which are relevant for this work.}. The ansatz reads
\begin{subequations}
\begin{align}
    \Lambda_i &= S\mathcal{A}_iS^{-1} + \frac{i}{g}S\partial_i S^{-1}\;,\label{ans1}\\
    \mathcal{A}_i &= (a-1)\frac{k}{g}\partial_i\phi\beta\cdot T\;,\\
    \psi_q&=h_{qp}ST_qS^{-1}\;,\\
    \psi_{\alpha/\bar{\alpha}} &= h_\alpha ST_{\alpha/\bar{\alpha}} S^{-1}\;,\\
    S&=e^{i\phi \beta\cdot T}\;,\;\beta=2N\Omega\label{ans2}\;.
\end{align}
\end{subequations}
Here, $\beta$ is the magnetic weight, and $\Omega$ is the highest weight of the representation of the static quarks. We used the notation $\beta\cdot T = \beta_q T_q$ and, for later convenience, we will assume $h_{-\alpha}=h_\alpha$ even though the profiles $h_\alpha$ were initially defined only for positive roots.  

These profile functions depend only on the cylindrical coordinate distance $\rho$ and must obey boundary conditions to reproduce smooth vortex configurations centered on the z-axis. At infinity, they must be such that the fields are in the vacuum manifold, that is
\begin{equation}
\label{boundaryconditions}
    \text{ When }\rho\rightarrow\infty\;,\; a\rightarrow1\;,\;h_{qp}\rightarrow v\delta_{qp}\;,\;h_\alpha\rightarrow v \;.
\end{equation}
Moreover, some of them must also obey regularity conditions in the vortex center $\rho=0$. The gauge profile $a(\rho)$ must vanish there to avoid a divergent magnetic field. As for the Higgs profiles, one must consider the behavior of the local frame
\begin{subequations}
    \begin{align}
        S T_q S^{-1}=&T_q\;,\\
        ST_{\alpha}S^{-1}=&\cos\left(\beta\cdot\alpha\,\varphi\right)T_{\alpha}-\sin\left(\beta\cdot\alpha\,\varphi\right)T_{\bar{\alpha}}\;,\\
        ST_{\bar{\alpha}}S^{-1}=&\sin\left(\beta\cdot\alpha\,\varphi\right)T_{\alpha}+\cos\left(\beta\cdot\alpha\,\varphi\right)T_{\bar{\alpha}}\;.
   \end{align}
\end{subequations}
This implies that if $\beta\cdot\alpha\ne0$, the field $\psi_\alpha$ is ill defined at $\rho=0$. This leads to the regularity condition at $ \rho=0$:
\begin{equation}
\label{regularityconditions}
        a=0\text{ and } h_\alpha=0 \text{ if } \beta\cdot\alpha \neq0\;.
\end{equation}

We will restrict the analysis to the $k-$Antisymmetric representations, as these are expected to give rise to the most stable confining strings in the asymptotic regime. Their highest weights are given by (see Appendix A for a very brief review of the weights of SU(N))
\begin{align}
    \Omega = \Omega^{(k)} = \sum_{i=1}^k\omega_i\;,
\end{align}
where $\omega_i$, $i=1,...,N$ are the weights of the fundamental representation. We must now show that the full equations of motion of the model can be reduced to a set of scalar ones for the profiles $a,h_{qp}, h_\alpha$. In this respect, let us initially investigate the implications for the gauge field equation
\begin{align}
    D_jF_{ij}=ig[\psi_A,D_i\psi_A]\;.
\end{align}
In Ref. \cite{OxmanSimoes2019}, this ansatz was shown to work for a potential that corresponds to the particular choice $m^2\ge 0, \kappa_d=0, \lambda_{i\neq 3}=0$ in the notation of the present paper. Since the equation for the gauge field is the same regardless of the potential, the same must hold for the general case. A nontrivial question is whether the equations for the Higgs fields close or not. These are
\begin{align}
    D^2\psi_A=\frac{\delta V_{\rm gen}}{\delta\psi_A}\;.\label{eom}
\end{align}
Using the commutation relations \eqref{commutation}, the left-hand side can be evaluated as
\begin{equation}
\label{lhsequationansatz}
    D^2\psi_q=\nabla^2 h_{qp} T_p\;,\;D^2\psi_{\alpha/\overline{\alpha}}=\left(\nabla^2 h_\alpha+(1-a)^2(\alpha\cdot \beta/\rho)^2\right) ST_{\alpha/\overline{\alpha}}S^{-1}\;.
\end{equation}

To present the results obtained for the right-hand side of \eqref{eom} (i.e. the forces), we define the following quantities
\begin{subequations}
\begin{align}
    F^{(2)-A}=\frac{1}{2}\frac{\delta V^{(2)}}{\delta \psi^A}\;&,\;F^{(4)-A}_i=\frac{1}{4}\frac{\delta V^{(4)}_i}{\delta \psi_A}\;,\;i=1,...,9\\F^{(3)-f-A}=\frac{1}{3}\frac{\delta V^{(3)-f}}{\delta \psi^A}\;&,\;F^{(3)-d-A}=\frac{1}{3}\frac{\delta V^{(3)-d}}{\delta \psi^A}\;.
\end{align} 
\end{subequations}
We shall start by analyzing the Cartan sector, i.e. when $A=q$. In light of eq. \eqref{lhsequationansatz}, the ansatz closes if the right-hand side is proportional to a combination of the Cartan generators. 

In the ansatz, the expressions for the lower-order forces, as well as those in the flavor-singlet category, are easier to obtain than the ones in the other categories.  For this reason, we simply present them below
\begin{subequations}
\label{ForceCartansector1}
    \begin{align}
    &F^{(2)-q}= h_{qp}T_p\makebox[.5cm],\makebox[.5cm]F^{(3)-f-q}=h_\alpha^2\alpha\vert_q\alpha\cdot T\makebox[.5cm],\makebox[.5cm]F^{(3)-d-q}=0\;,\\
    &F^{(4)-q}_1=\left(\rm{Tr}(\mathbb{H}^T\mathbb{H})+2h_\alpha^2\right)h_{qp}T_p\makebox[.4cm],\makebox[.4cm]
    F^{(4)-q}_2=h_{ql}h_{pl}h_{pm}T_m\makebox[.4cm],\makebox[.4cm]    F^{(4)-q}_3=2h_{\alpha}^2h_{qp}\alpha\vert_p\alpha\cdot T\;,\\
    &F^{(4)-q}_4=(4N)^2h_{qp}\omega_j\vert_p(\omega_i\cdot \mathbb{H}^T \mathbb{H}.\omega_i)(\omega_i\cdot\omega_j)\omega_j\cdot T+8Nh_\alpha^2(\tilde{\alpha}\cdot\omega_i)h_{qp}\omega_i\vert_p\omega_i\cdot T\;,   
    \end{align}
\end{subequations}
where $\mathbb{H}\vert_{qp}=h_{qp}$,  $i$ and $j$ are summed from $1$ to $N$ and $\alpha$ is summed over the positive roots. 

When it comes to $F^{(4)-q}_{i\ge5}$, the calculations are more subtle. For illustrative purposes, we will show the main steps to compute $F^{(4)}_6$ since it represents well the overall complexity. Recalling eq. \eqref{V6}, we have
\begin{subequations}
    \begin{align}
    F^{(4)-A}_6 &=f_{ABE}f_{CDE}\psi_B\wedge(\psi_C\wedge\psi_D)\;.
    \end{align}
\end{subequations}
The first step is to consider all the non-vanishing possibilities for the indices $B$, $C$, and $D$ given that $A=q$, i.e. 
\begin{eqnarray}
   \label{F6qexample}
   F_6^{(4)-q} &=&2f_{q\alpha\overline{\alpha}}f_{p\alpha\overline{\alpha}}\psi_\alpha\wedge\left(\psi_p\wedge\psi_\alpha\right)+2f_{q\overline{\alpha}\alpha}f_{p\overline{\alpha}\alpha}\psi_{\overline{\alpha}}\wedge\left(\psi_p\wedge\psi_{\overline{\alpha}} \right)+f_{q\alpha\overline{\alpha}}f_{\gamma\eta\overline{\alpha}}\psi_\alpha\wedge\left(\psi_\gamma\wedge\psi_{\eta}\right)\notag\\
    &&+f_{q\alpha\overline{\alpha}}f_{\bar{\gamma}\bar{\eta}\overline{\alpha}}\psi_\alpha\wedge\left(\psi_{\bar{\gamma}}\wedge\psi_{\bar{\eta}}\right)+2f_{q\overline{\alpha}\alpha}f_{\gamma\bar{\eta}\alpha}\psi_{\overline{\alpha}}\wedge\left(\psi_\gamma\wedge\psi_{\bar{\eta}}\right)\;,
\end{eqnarray}
where the sum in $\alpha$ is to be performed over all positive roots and in $\gamma$ and $\eta$ over all positive roots as long as $\alpha\ne\eta\ne\gamma\ne\alpha$. Using the properties of the structure constants, we can reduce eq. \eqref{F6qexample} to a sum of two terms. The first one can be readily evaluated using eqs. \eqref{commutation} and \eqref{fCartan}:
\begin{equation}
4f_{q\alpha\overline{\alpha}}f_{p\alpha\overline{\alpha}}\psi_\alpha\wedge\left(\psi_p\wedge\psi_\alpha\right)=4\alpha\vert_q\alpha\vert_p h_{pl} h_\alpha^2\alpha_l\alpha_r T_r=4h_\alpha^2\left(\alpha\cdot \mathbb{H}\cdot \alpha\right)\alpha\vert_q\alpha\cdot T\;.
\end{equation}
The second one reads
\begin{equation}
4f_{q\alpha\overline{\alpha}}f_{\gamma\eta\overline{\alpha}}\psi_\alpha\wedge\left(\psi_\gamma\wedge\psi_{\eta}\right)=4h_\alpha h_\gamma h_\eta\alpha\vert_q f_{\gamma\eta\overline{\alpha}}f_{\gamma\eta\overline{\delta}}S\left( T_\alpha\wedge T_{\overline{\delta}}\right) S^{-1}\;,
\end{equation}
where $\alpha$, $\gamma$, $\eta$ and $\delta$ are summed over the positive roots. To simplify this expression further, eq. \eqref{froots2} plays a crucial role. First, notice that the above expression vanishes unless $\alpha=\delta$ since both $\delta$ and $\alpha$ are positive roots. The positivity is important here because, otherwise, eq.  \eqref{froots2} would allow other possibilities like, for example, $\alpha=\gamma-\eta=-\delta$. Also, using eq. \eqref{Nproperty}, we find
\begin{equation}
    f_{\gamma\eta\overline{\alpha}}^2 = \frac{1}{2}N_{\gamma,\eta}^2(\delta_{\alpha,\gamma+\eta}+\delta_{\alpha,-\gamma-\eta})+\frac{1}{2}N_{\gamma,-\eta}^2(\delta_{\alpha,\gamma-\eta}+\delta_{\alpha,\eta-\gamma}) = \frac{1}{2}N_{\gamma,\eta}^2\delta_{\alpha,\gamma+\eta}\;,
\end{equation}
where we changed the sum on $\gamma$ and $\eta$ in the last equality to be overall roots, positive and negative, as long as $\gamma\ne\eta\ne\alpha\ne\gamma$. This implies
\begin{eqnarray}
    4f_{q\alpha\overline{\alpha}}f_{\gamma\eta\overline{\alpha}}\psi_\alpha\wedge\left(\psi_\gamma\wedge\psi_{\eta}\right)=2N^2_{\alpha,\gamma}h_\alpha h_\gamma h_{\alpha+\gamma}\alpha\vert_q\alpha\vert_pT_p\;.
\end{eqnarray}

Additionally, provided one keeps in mind the properties of the symmetric constants (see eqs. \eqref{anticommutation} and \eqref{dconstants}), the previous remarks can be readily applied to all of the other forces over the field $\psi_q$. After doing so, the expressions for f-adjoint forces are
\begin{subequations}
\label{ForceCartansector2}
\begin{align}
    F^{(4)-q}_5&=2h_\alpha^2\alpha\vert_q \alpha\cdot\mathbb{H}\cdot T\;,\\ 
     F^{(4)-q}_6&=4h_\alpha^2\alpha\vert_q\left(\alpha\cdot \mathbb{H}\cdot\alpha\right)\alpha\cdot T+2N^2_{\alpha,\gamma}h_{\alpha} h_{\gamma}h_{\alpha+\gamma}\alpha\vert_q\alpha\cdot T\;.\label{ForceCartansector2:2}
     \end{align}
\end{subequations}

In the d-adjoint case, the forces after applying the ansatz become
\begin{subequations}
\label{ForceCartansector3}
    \begin{align}
     F^{(4)-q}_{7}=&(4N)^2\omega_i\vert_q(\omega_i\cdot\omega_j)(\omega_j.\mathbb{H}\mathbb{H}^T\cdot\omega_j)\left(\omega_i\cdot \mathbb{H}\cdot T\right) +8Nh_\alpha^2\omega_i\vert_q\left(\omega_i\cdot\tilde{\alpha}\right)\left(\omega_i\cdot \mathbb{H}\cdot T\right)\;,\\
     F^{(4)-q}_{8}=&(4N)^4\left(\omega_i\cdot \mathbb{H}\cdot\omega_b\right)\left(\omega_i\cdot\omega_j\right)\left(\omega_j\cdot \mathbb{H}\cdot\omega_a\right)^2\left(\omega_a\cdot\omega_b\right)\omega_i\vert_q\omega_b\cdot T+4h_\alpha^2\tilde{\alpha}\vert_q\left(\tilde{\alpha}\cdot \mathbb{H}\cdot\tilde{\alpha}\right)\tilde{\alpha}\cdot T\notag\\
     &+2(4N)^2h_\alpha^2(\omega_i\cdot \mathbb{H}\cdot\omega_j)(\tilde{\alpha}\cdot\omega_i)(\tilde{\alpha}\cdot\omega_j)\omega_i\vert_q\omega_j\cdot T+2M^2_{\alpha,\gamma}h_{\alpha}h_{\gamma}h_{\alpha+\gamma} \tilde{\alpha}_q\tilde{\alpha}\cdot T\;,
    \end{align}
\end{subequations}

while the mixed adjoint one turns out to be
\begin{eqnarray}
\label{ForceCartansector4}
    F^{(4)-q}_{9}&=&2\alpha\vert_q\left(\tilde{\alpha}\cdot \mathbb{H}\cdot \tilde{\alpha}\right)h_\alpha^2\alpha\cdot T + N^2_{\alpha\gamma}h_{\alpha}h_{\gamma}h_{\alpha+\gamma}\left(\alpha\vert_q\alpha\vert_p+\tilde{\alpha}_q\tilde{\alpha}_p\right) T_p\notag\\
     &&+(4N)^2\left(\omega_i\cdot \mathbb{H}\cdot \omega_j\right)(\omega_i\cdot \alpha)(\omega_j\cdot \alpha)h_{\alpha}^2\omega_i\vert_q\omega_j\cdot T+2(\alpha\cdot \mathbb{H}\cdot \alpha)h_{\alpha}^2\tilde{\alpha}\vert_q\tilde{\alpha}\cdot T\;.
\end{eqnarray}
Again, in the above expressions,  $\alpha$ must be summed over the positive roots while $\gamma$ must be summed over all of the roots, provided $\gamma\ne \pm\alpha$. The indices $i$, $j$, $a$, and $b$ label the weights of the fundamental representation, thus ranging from $1$ to $N$. Because of the terms involving the symmetric constants $d_{ABC}$, we defined a vector $\tilde{\alpha}=\omega_a+\omega_b$ for each root $\alpha=\omega_a-\omega_b$ (for details, see Appendix B).

In the root sector, i.e. setting $A=\alpha$, we have
\begin{eqnarray}
    F_6^{(4)-\alpha}&=&2f_{\alpha q \bar{\alpha}}f_{p\alpha\bar{\alpha}}\psi_{q}\wedge\left(\psi_{p}\wedge\psi_{\alpha}\right)+f_{\alpha q\bar{\alpha}}f_{\gamma\eta\bar{\alpha}}\psi_{q}\wedge\left(\psi_{\gamma}\wedge\psi_{\eta}\right)+f_{\alpha q\bar{\alpha}}f_{\bar{\gamma}\bar{\eta}\bar{\alpha}}\psi_{q}\wedge\left(\psi_{\bar{\gamma}}\wedge\psi_{\bar{\eta}}\right)\notag\\
    &&+2f_{\alpha\bar{\alpha}q}f_{\alpha'\bar{\alpha}'q}\psi_{\bar{\alpha}}\wedge\left(\psi_{\alpha'}\wedge\psi_{\bar{\alpha}'}\right)+f_{\alpha\gamma\bar{\delta}}f_{\varepsilon\xi\bar{\delta}}\psi_{\gamma}\wedge\left(\psi_{\varepsilon}\wedge\psi_{\xi}\right)+f_{\alpha\gamma\bar{\delta}}f_{\bar{\varepsilon}\bar{\xi}\bar{\delta}}\psi_{\gamma}\wedge\left(\psi_{\bar{\varepsilon}}\wedge\psi_{\bar{\xi}}\right)\notag\\
    &&+2f_{\alpha\bar{\gamma}\delta}f_{\varepsilon\bar{\xi}\delta}\psi_{\gamma}\wedge\left(\psi_{\varepsilon}\wedge\psi_{\bar{\xi}}\right)\;.
\end{eqnarray}
Here, there is no sum over $\alpha$ and there are sums over all $\alpha'>0$ and all roots $\gamma,\eta,\varepsilon,\xi$, both positive and negative. The above expression brings two notable differences when compared to its Cartan sector counterpart. First, there are terms with summations over 5 positive roots, instead of just 3. These terms can be treated just like before but with the use of eq. \eqref{fconstants} twice. Second, a completely new kind of term shows up, namely
\begin{equation}
    2f_{\alpha\bar{\alpha}q}f_{\alpha'\bar{\alpha}'q}\psi_{\bar{\alpha}}\wedge\left(\psi_{\alpha'}\wedge\psi_{\bar{\alpha}'}\right)=(\alpha\cdot\alpha')^2h_{\alpha'}^2h_\alpha ST_\alpha S^{-1}\;.
\end{equation}
Here, the $\alpha$ and $\alpha'$ are positive roots and can even be the same, which is why $\alpha'$ was used, instead of a different greek letter. Other than that, the steps to compute the forces in the root sector are similar to what was done before and we start again by exhibiting first the result for the lower order forces and the flavor-singlet ones
\begin{subequations}
\label{Forcerootsector1}
    \begin{align}
    F^{(2)-\alpha}=&h_\alpha S T_\alpha S^{-1}\;,\\
    F^{(3)-f-\alpha}=&2\left(\alpha\cdot\mathbb{H}\cdot\alpha\right)h_{\alpha}ST_{\alpha}S^{-1}+N^2_{\alpha,\gamma}h_{\gamma}h_{\alpha+\gamma}ST_{\alpha}S^{-1}\;,\\
    F^{(3)-d-\alpha}=&2\left(\tilde{\alpha}\cdot\mathbb{H}\cdot\tilde{\alpha}\right)h_{\alpha}ST_{\alpha}S^{-1}+N^2_{\alpha,\gamma}h_{\gamma}h_{\alpha+\gamma}ST_{\alpha}S^{-1}\;,\\
        F^{(4)-\alpha}_1=&\big(\rm{Tr}(\mathbb{H}^T\mathbb{H})+2h_{\alpha'}^2\big)h_{\alpha}ST_{\alpha}S^{-1}\;,\\
     F^{(4)-\alpha}_2=&h_{\alpha}^3ST_{\alpha}S^{-1}\;,\\
     F^{(4)-\alpha}_3=&\big(\alpha\cdot \mathbb{H}^T\mathbb{H}\cdot\alpha+N_{\alpha,\gamma}^2h_{\gamma}^2+|\alpha|^2h_{\alpha}^2\big)h_{\alpha} ST_{\alpha}S^{-1}\;,\\
     F^{(4)-\alpha}_4=&\left(4N(\omega_i\cdot\mathbb{H}^T\mathbb{H}\cdot\omega_i)(\omega_i\cdot\tilde{\alpha})+2(\tilde{\alpha}\cdot\tilde{\alpha}')h_{\alpha'}^2\right)h_{\alpha}ST_{\alpha}S^{-1}\;.
    \end{align}
\end{subequations}
Then, we move to the f-adjoint set
\begin{subequations}
\label{Forcerootsector2}
    \begin{align}
        F^{(4)-\alpha}_5=&\left(\alpha\cdot\mathbb{H} \mathbb{H}^T\cdot\alpha+N^2_{\alpha,\gamma}h_\gamma^2+|\alpha|^2h_{\alpha}^2\right)h_{\alpha}ST_{\alpha}S^{-1}\;,\\
     F^{(4)-\alpha}_6=&\left(2(\alpha\cdot\mathbb{H}\cdot\alpha)^2+2\left(\alpha\cdot\alpha '\right)^2h_{\alpha'}^2 \right)h_{\alpha}ST_{\alpha}S^{-1}+N_{\alpha,\gamma }^2\big(\alpha\cdot\mathbb{H}\cdot\alpha\big)h_{\gamma}h_{\alpha+\gamma}ST_{\alpha}S^{-1}\notag\\
     & +\big(\gamma\cdot\mathbb{H}\cdot\gamma\big) N_{\alpha ,\gamma}^2h_{\gamma}h_{\alpha+\gamma}ST_{\alpha}S^{-1}+N_{\alpha,\gamma}^2N_{\gamma,\eta}^2h_{\eta}h_{\gamma+\eta}h_{\alpha +\gamma}ST_{\alpha}S^{-1} \;,
    \end{align}
\end{subequations}
while the d-ajoint forces read
\begin{subequations}
\label{Forcerootsector3}
\begin{align}
     F^{(4)-\alpha}_{7}=&\big(4N(\tilde{\alpha}\cdot\omega_i)(\omega_i\cdot\mathbb{H}\mathbb{H}^T\cdot\omega_i)+2(\tilde{\alpha}\cdot\tilde{\alpha}')h_{\alpha'}^2\big)h_{\alpha}ST_{\alpha}S^{-1}\;,\\
     F^{(4)-\alpha}_{8}=&2\left(\tilde{\alpha}\cdot\mathbb{H}\cdot\tilde{\alpha}\right)^2h_{\alpha}ST_{\alpha}S^{-1}+(4N)^2\left(\tilde{\alpha}\cdot\omega_i\right)\left(\tilde{\alpha}\cdot\omega_j\right)\left(\omega_i\cdot\mathbb{H}\cdot\omega_j\right)^2h_{\alpha}ST_{\alpha}S^{-1}\notag\\
     &+M^2_{\alpha,\gamma}\left(\tilde{\alpha}\cdot\mathbb{H}\cdot\tilde{\alpha}\right)h_{\gamma}h_{\alpha+\gamma}ST_{\alpha}S^{-1}+2\left(\tilde{\gamma}\cdot\mathbb{H}\cdot\tilde{\gamma}\right)M^2_{\alpha,\gamma}h_{\alpha+\gamma}h_{\gamma}ST_{\alpha}S^{-1}\notag\\
     &+2\left(\tilde{\alpha}\cdot\tilde{\alpha}'\right)^2h_{\alpha}h_{\alpha'}^2ST_{\alpha}S^{-1}+M^2_{\alpha,\gamma}M^2_{\gamma,\eta}h_{\alpha+\gamma}h_{\eta}h_{\gamma+\eta}ST_{\alpha}S^{-1}\;.
    \end{align}
\end{subequations}
Finally, the mixed-adjoint force reads
\begin{align}
\label{Forcerootsector4}
    F^{(4)-\alpha}_{9}=&2\left(\alpha\cdot\mathbb{H}\cdot\alpha\right)\left(\tilde{\alpha}\cdot\mathbb{H}\cdot\tilde{\alpha}\right)h_{\alpha}ST_{\alpha}S^{-1}+\left(\frac{1}{2}\alpha\cdot\mathbb{H}\cdot\alpha+\frac{1}{2}\tilde{\alpha}\cdot\mathbb{H}\cdot\tilde{\alpha}\right)M^2_{\alpha,\gamma}h_{\gamma}h_{\alpha+\gamma}ST_{\alpha}S^{-1}\notag\\
     &+8N^2\left(\alpha\cdot\omega_i\right)\left(\alpha\cdot\omega_j\right)\left(\omega_i\cdot\mathbb{H}\cdot\omega_j\right)^2h_{\alpha}ST_{\alpha}S^{-1}+((\alpha\cdot\tilde{\alpha}')^2+(\alpha'\cdot\tilde{\alpha})^2)h_{\alpha'}^2h_{\alpha}ST_{\alpha}S^{-1}\notag\\
     &+\big(\gamma \cdot\mathbb{H}\cdot\gamma+\tilde{\gamma}\cdot\mathbb{H}\cdot\tilde{\gamma}\big)N^2_{\alpha,\gamma}h_{\gamma}h_{\alpha+\gamma}ST_{\alpha}S^{-1}+N_{\gamma ,\eta }^2 N_{\alpha ,\gamma }^2 h_{\eta } h_{\gamma +\eta }
    h_{\alpha +\gamma }ST_{\alpha}S^{-1} \;.
\end{align}
The summations are over $i,j=1,...,N$, $\alpha'>0$, including $\alpha'=\alpha$, positive and negative $\gamma,\eta$, excluding $\gamma=\pm\alpha\ne\eta\ne\gamma$. The expression for the forces with index $\bar{\alpha}$ are the same after replacing barred roots indices with unbarred ones and vice-versa. 

Since we showed that both sides of eq. \eqref{eom} point in the same direction in the Lie algebra, then our ansatz closes and we are left with scalar equations for the profiles $a$, $h_{qp}$, and $h_{\alpha}$.

\section{Abelianization  and the  asymptotic Casimir law}
\label{AbelianizationPoint}

 Here, we shall discuss the energy scaling of the solution with $N-$ality $k$. A good starting point is to review some facts regarding the particular case of the model given by eq. \eqref{higgs}, which was studied in Ref. \cite{OxmanSimoes2019}. It has a special point in parameter space, $m^2=0$, where all the profiles $h_\alpha$ with $\alpha\cdot\beta=0$ freeze at their vacuum value $v$. As for the other profiles, a collective behavior was shown to take place, i.e. $h_\alpha=h$ for all $\alpha$ with $\alpha\cdot\beta\ne0$. In this case, the model can be said to be Abelianized as the equations satisfied by the profiles $a,h$ are those of the Ginzburg-Landau model, which gives rise to the well-known Nielsen-Olesen vortex. The string tension (energy per unit length) in this particular case is
\begin{align}
    \sigma_{\rm particular}=k(N-k)\int d^2 x\, \left(\frac{|\nabla a|^2}{\rho^2g^2}+\frac{h^2(1-a)^2}{\rho^2}+|\nabla h|^2\right)+V_{\rm particular}\;.
\end{align}
Here, we can use Derricks's theorem in two dimensions, which states that the kinetic energy of the gauge field is equal to the potential energy of the Higgs fields, thus implying 
\begin{align}
    \sigma_{\rm particular}=k(N-k)\int d^2 x\, \left(2\frac{|\nabla a|^2}{\rho^2g^2}+\frac{h^2(1-a)^2}{\rho^2}+|\nabla h|^2\right)\;.
\end{align}
This string tension scales with the quadratic Casimir of the $k-$Antisymmetric representation, in accordance with the Casimir law 
\begin{align}
    \frac{\sigma_k}{\sigma_1}=\frac{k(N-k)}{N-1}
\end{align}
approximately observed in the lattice. The above derivation makes it clear that a set of ingredients for such a law is the existence of a region in parameter space for which the following conditions are met
\begin{enumerate}
    \item  $h_\alpha=v,\,\forall \alpha |\alpha\cdot\beta=0 $,
\item $h_\alpha=h,\,\forall \alpha |\alpha\cdot\beta=1 $,
\item $h$ must be independent of $k$.
\end{enumerate}
Keep in mind that only two values for $\alpha\cdot\beta$ are being considered since the weight in eq. \eqref{ans2} are those of the $k-$Antisymmetric representations.

In the following, we will analyze the existence of such a region for the model of eq. \eqref{gen_model}. Then, let us assume
\begin{eqnarray}
     &h_{qp}=v \delta_{qp}\;,\nonumber\\&
     \;h_{\alpha}=\begin{cases}
         v \text{ if }\alpha\cdot\beta=0\\
         h\text{ if }\alpha\cdot\beta=1
     \end{cases}\;
\end{eqnarray}
and evaluate the force expressions \eqref{ForceCartansector1}, \eqref{ForceCartansector2}-\eqref{ForceCartansector4}, and \eqref{Forcerootsector1}-\eqref{Forcerootsector4}. 

Once again, we will illustrate the calculations using $F_6$ as an example. In the Cartan sector,
\begin{equation}
\label{F6qevaluationstart}
    F^q_6=4h_\alpha^2\alpha\vert_q\left(\alpha\cdot \mathbb{H}\cdot\alpha\right)\alpha\cdot T+2N^2_{\alpha,\gamma}h_{\alpha} h_{\gamma}h_{\alpha+\gamma}\alpha\vert_q\alpha\cdot T
\end{equation}

To evaluate these terms, we first need to write down explicitly, for each $k$, which positive root yields the value $0$ or $1$ for the product $\alpha\cdot\beta$. If we express the roots as differences of weights of the fundamental representation, i. e. $\alpha=\alpha_{ij}=\omega_i-\omega_j$, we have
\begin{eqnarray}
\label{roottypes}
\alpha_{ij}\cdot\beta=\begin{cases}
    0\text{, if } i,j=1,...,k\text{ or } i,j=k+1,...,N\;,\\
   1\text{, if } i=1,...,k\text{ and }j=k+1,...,N\;,
\end{cases}    
\end{eqnarray}
and the positivity of the root is guaranteed by $i<j$. Now, we can define the matrices
\begin{subequations}
    \begin{align}
A_1\vert_{qp}^{(k)}=&\sum\limits_{i=1}^{k}\sum\limits_{j=k+1}^N\alpha_{ij}\vert_q\alpha_{ij}\vert_p\;,\\
A_{0}\vert_{qp}^{(k)}=&\sum\limits_{i=k+1}^{N}\sum\limits_{j=i+1}^N\alpha_{ij}\vert_q\alpha_{ij}\vert_p\,,\\
\tilde{A}_{0}\vert_{qp}^{(k)}=&\sum\limits_{i=1}^{k}\sum\limits_{j=i+1}^{k}\alpha_{ij}\vert_q\alpha_{ij}\vert_p\;,
    \end{align}
\end{subequations}
which sum up to half the identity matrix. Then, noticing every root has length equal to $1/\sqrt{N}$, the first term in eq. \eqref{F6qevaluationstart} is
\begin{eqnarray}
    4h_\alpha^2\alpha\vert_q\left(\alpha\cdot \mathbb{H}\cdot\alpha\right)\alpha\cdot T &=&\frac{4}{N}v^3\sum\limits_{^{\;\;\alpha>0}_{\alpha\cdot\beta^k=0}}\alpha\vert_q\alpha\cdot T + \frac{4}{N}vh^2\sum\limits_{^{\;\;\alpha>0}_{\alpha\cdot\beta^k\ne0}}\alpha\vert_q\alpha\cdot T \;,\notag\\
    &=& \frac{2}{N}v^3T_q+ \frac{4}{N}v(h^2-v^2)A^k_{qp} T_p\;.
\end{eqnarray}
To evaluate the second term in eq. \eqref{F6qevaluationstart}, we need to split the sum over $\alpha$ and $\gamma$ into different cases to take into account all the possibilities for the profiles $h_\alpha h_\gamma h_{\alpha+\gamma}$. However, the matrix part of the term depends only on $\alpha$ and not $\gamma$. This means that in each case the sum over $\gamma$ only contributes with a numerical factor. The following table summarizes the different cases. The first column shows how the roots $\alpha$ and $\gamma$ must match to ensure $\alpha+\gamma$ is a valid root. The second column shows the range of the indices $i,j,k$ that label the roots. The third one show the associated profiles and the last one shows the multiplicity for each case, i.e. how many $\gamma$ possibilities are there for each $\alpha$.

\begin{center}
\begin{tabular}{cccc}
Root type & Indices range &Profiles $(h_\alpha,\; h_\gamma,\; h_{\alpha+\gamma})$& Multiplicity\\
$\alpha_{ij}\;,\;\gamma_{jl}$ & $i\le k\;,\;j>k\;,\;l\le k$& $(h,h,\tilde{h}_0)$ & $(k-1)$ \\
$\alpha_{ij}\;,\; \gamma_{li}$ & $i\le k\;,\;j>k\;,\;l> k$& $(h,h,h_0)$ & $(N-k-1)$ \\
$\alpha_{ij}\;,\;\gamma_{li}$ & $i\le k\;,\;j>k\;,\;l\le k$& $(h,\tilde{h}_0,h)$ & $(k-1)$ \\
$\alpha_{ij}\;,\;\gamma_{jl}$ & $i\le k\;,\;j>k\;,\;l> k$& $(h,h_0,h)$ & $(N-k-1)$ \\
$\alpha_{ij}\;,\; \gamma_{li}$ or $\gamma_{jl}$ & $i\le k\;,\;j\le k\;,\;l>k$& $(\tilde{h}_0,h,h)$ & $2(N-k)$\\
$\alpha_{ij}\;,\;\gamma_{li}$ or $\gamma_{jl}$ & $i\le k\;,\;j\le k\;,\;l\le k$& $(\tilde{h}_0,\tilde{h}_0,\tilde{h}_0)$ & $2(k-2)$\\
$\alpha_{ij}\;,\;\gamma_{li}$ or $\gamma_{jl}$ & $i> k\;,\;j> k\;,\;l\le k$& $(h_0,h,h)$ & $2k$\\
$\alpha_{ij}\;,\;\gamma_{li}$ or $\gamma_{jl}$ & $i> k\;,\;j>k\;,\;l>k$& $(h_0,h_0,h_0)$ & $2(N-k-2)$
\end{tabular}
\end{center}
Then, the second term reads
\begin{eqnarray}
    2N^2_{\alpha,\gamma}h_{\alpha} h_{\gamma}h_{\alpha+\gamma}\alpha\vert_q\alpha\cdot T&=&\frac{2(h^2-v^2)v}{N}\left((N-2)A\vert^{(k)}_{qp}T_p+(N-k)\tilde{A}_0\vert^{(k)}_{qp}T_p+kA_0\vert^{(k)}_{qp}T_p\right)\notag\\
    &&+\frac{2v^3(N-2)}{N}\left(\tilde{A}_0\vert^{(k)}_{qp}T_p+A_0\vert^{(k)}_{qp}T_p+A\vert^{(k)}_{qp}T_p\right)  \;.
\end{eqnarray}
We can eliminate  $A_0^{(k)}$ and $\tilde{A}_0^{(k)}$ by using the identities
\begin{eqnarray}
    A_{0}\vert^{(k)}_{qp}&=&\frac{2N-k}{4N}\delta_{qp}-\frac{N}{2}\beta\vert_{q}\beta\vert_{p}-\frac{A\vert^{(k)}_{qp}}{2}\;,\\
     \tilde{A}_0\vert^{(k)}_{qp}&=&\frac{k}{4N}\delta_{qp}+\frac{N}{2}\beta\vert_{q}\beta\vert_{p}-\frac{A\vert^{(k)}_{qp}}{2}\;,
\end{eqnarray}
which leads to
\begin{eqnarray}
    2N^2_{\alpha,\gamma}h_{\alpha} h_{\gamma}h_{\alpha+\gamma}\alpha\vert_q\alpha\cdot T&=&\frac{v^3(N-2)}{N}T_q+
    \frac{k(3N-2k)}{2N^2}v(h^2-v^2)T_q\notag\\
    &&+\frac{N-4}{N}v(h^2-v^2)A\vert^{(k)}_{qp}T_p +(N-2k)v(h^2-v^2)\beta\vert_{q}\beta\cdot T 
\end{eqnarray}
A similar analysis can be carried out for all of the other forces. Just as before, we start by showing the result for the lower order and flavor-singlet interactions
\begin{subequations}
\label{SingletCartanansatzForces}
\begin{align}
    F^{(2)-q}=&v T_{q}\;,\\
    F^{(3)-f-q}=&v^2T_{q}+2(h^2-v^2)A\vert^{(k)}_{qp}T_p\;,\\
    F^{(3)-d-q}=&\frac{N^2-4}{N^2}v^2T_{q}+\frac{2k}{N}(h^2-v^2)T_q-2(h^2-v^2)A\vert^{(k)}_{qp}T_p\notag\\
    &+4(h^2-v^2)(N-2k)\beta\vert_{q}\beta\cdot T\;,\\
    F^{(4)-q}_{1}=&(N^2-1)v^3T_q+2k(N-k)v(h^2-v^2)T_q\;,\\
    F^{(4)-q}_{2}=&v^3T_q\;,\\
    F^{(4)-q}_{3}=&v^3T_q+2v(h^2-v^2)A\vert^{(k)}_{qp}T_p\;,\\
    F^{(4)-q}_{4}=&-\frac{2k(N-2k)}{N^2}v(h^2-v^2)T_q+4(N-2k)v(h^2-v^2)\beta\vert_{q}\beta\cdot T\;.
\end{align}
\end{subequations}
Next, we present the result for the f-adjoint set of interactions 
\begin{subequations}
\label{fadjointCartanansatzForces}
    \begin{align}
    F^{(4)-q}_{5}=&v^3T_q+2v(h^2-v^2)A\vert^{(k)}_{qp}T_p\;,\\
    F^{(4)-q}_{6}=&v^3T_q+\frac{k(3N-2k)}{2N^2}v(h^2-v^2)T_q+v(h^2-v^2)A\vert^{(k)}_{qp}T_p\notag\\
    &+(N-2k)v(h^2-v^2)\beta\vert_{q}\beta\cdot T\;,
    \end{align}
\end{subequations}
then the d-adjoint forces
\begin{subequations}
\label{dadjointCartanansatzForces}
    \begin{align}
    F^{(4)-q}_{7}&=-\frac{2k(N-2k)}{N^2}v(h^2-v^2)T_q+4(N-2k)v(h^2-v^2)\beta\vert_{q}\beta\cdot T\;,\\
    F^{(4)-q}_{8}&=\left(\frac{N^2-4}{N^2}\right)^2v^3T_q-\frac{k(24N-5N^3)+k^2(16+2N^2)}{2N^4}v(h^2-v^2)T_q\notag\\
    &-\frac{N^2-12}{N^2}v(h^2-v^2)A\vert^{(k)}_{qp}T_p+\frac{(3N^2-40)(N-2k)}{N^2}v(h^2-v^2)\beta\vert_{q}\beta\cdot T\;,
    \end{align}
\end{subequations}
and the mixed-adjoint one
\begin{eqnarray}
\label{mixedCartanansatzForces}
    F^{(4)-q}_{9}&=&\frac{N^2-4}{N^2}v^3T_q+\frac{k(2N-k)}{N^2}v(h^2-v^2)T_q\notag\\
    &&-\frac{6}{N^2}v(h^2-v^2)A\vert^{(k)}_{qp}T_p+2(N-2k)v(h^2-v^2)\beta\vert_{q}\beta\cdot T\;.
\end{eqnarray}
Notice that these forces are combinations of the following five expressions: $T_q$, $k(h^2-v^2)T_q$, $k^2(h^2-v^2)T_q$, $(h^2-v^2)A\vert^{(k)}_{qp}T_p$, and $(N-2k)(h^2-v^2)\beta\vert_{q}\beta\cdot T$. As discussed before, if the profiles $h_{qp}$ were to freeze at their vacuum values, a Casimir scaling can be found. Because of eq. \eqref{lhsequationansatz}, this means that the total force on the fields $\psi_q$ must vanish, which implies equating the total coefficients of each piece to $0$. Doing so for the coefficient of $T_q$ leads to
\begin{eqnarray}
    0&=&m^2+\kappa_f v+ \frac{N^2-4}{N^2}\kappa_d v+ (N^2-1)\lambda_1 v^2+\lambda_2v^2+\lambda_3v^2\notag\\
    &&+\lambda_5v^2+\lambda_6v^2+\left(\frac{N^2-4}{N^2}\right)^2 \lambda_{8} v^2+ \frac{N^2-4}{N^2} \lambda_{9} v^2\;,\label{Vacuumvcondition}
\end{eqnarray}
which is actually what defines the value of $v\ne0$ that minimizes the potential. As for the other pieces, they yield a set of four conditions, out of which only three are independent:
\begin{subequations}
\label{cartanconditions}
\begin{align}
    &2\kappa_d+2N^2\lambda_1v-2\lambda_4v+\frac{3}{2}\lambda_6v-2\lambda_{7}v+\frac{5N^2-24}{2N^2}\lambda_{8}v+2\lambda_{9}v=0\label{firstcartancondition}\\
    &2N^2\lambda_1-4\lambda_4+\lambda_6-4\lambda_{7}+\frac{N^2+8}{N^2}\lambda_{8}+\lambda_{9}=0\label{secondcartancondition}\\
    &2\kappa_f-2\kappa_d+2\lambda_3v+2\lambda_5v+\lambda_6v-\frac{N^2-12}{N^2}\lambda_{8}v-\frac{6}{N^2}\lambda_{9}v=0\label{thirdcartancondition}
\end{align}
\end{subequations}
In the root sector, $F_\alpha$ was shown to be proportional to $ST_\alpha S^{-1}$ in eqs. \eqref{Forcerootsector1}-\eqref{Forcerootsector4}, but the expression for the forces changes depending on the type of root as defined in eq. \eqref{roottypes}. If the roots $\alpha=\omega_i-\omega_j$ are perpendicular to the magnetic weight and $i,j>k$, the lower order and flavor-singlet forces are\footnote{These expressions are the same for all values of $i$ and $j$, provided $i,j>k$. Additionally, when $i,j<k$, the forces can be obtained from the former case by a simple change $k\rightarrow N-k$. As these properties are true for all of the forces acting on $\psi_{\alpha/\bar{\alpha}}$, we will omit the case $i,j<k$.}
\begin{subequations}
\label{Root0singletansatzForces}
\begin{align}
    \langle F^{(2)-\alpha},ST_{\alpha}S^{-1}\rangle&=v\;,\\
    \langle F^{(3)-f-\alpha},ST_{\alpha}S^{-1}\rangle&=v^2+\frac{k}{N}(h^2-v^2)\;,\\
    \langle F^{(3)-d-\alpha},ST_{\alpha}S^{-1}\rangle&= \frac{N^2-4}{N^2}v^2 + \frac{k}{N}(h^2-v^2)\;,\\
    \langle F^{\alpha}_{1},ST_{\alpha}S^{-1}\rangle&=(N^2-1)v^3 + 2k(N-k)v(h^2-v^2)\;,\\
    \langle F^{\alpha}_{2},ST_{\alpha}S^{-1}\rangle&=v^3\\
    \langle F^{\alpha}_{3},ST_{\alpha}S^{-1}\rangle&=v^3 + \frac{kv(h^2-v^2)}{N}\;,\\
    \langle F^{\alpha}_{4},ST_{\alpha}S^{-1}\rangle&=-2\frac{k(N-2k)}{N^2}v(h^2-v^2)\;.
\end{align}
\end{subequations}
The forces in the f-adjoint set are
\begin{subequations}
\label{Root0fadjointtansatzForces}
    \begin{align}
    \langle F^{\alpha}_{5},ST_{\alpha}S^{-1}\rangle&=v^3+\frac{k}{N}v(h^2-v^2)\;,\\
    \langle F^{\alpha}_{6},ST_{\alpha}S^{-1}\rangle&=v^3+\frac{k(2N-k)}{N^2}v(h^2-v^2)\;,
    \end{align}
\end{subequations}
while those in the d-adjoint set are
\begin{subequations}
\label{Root0dadjointansatzForces}
    \begin{align}
    \langle F^{\alpha}_{7},ST_{\alpha}S^{-1}\rangle&=-2\frac{k(N-2k)}{N^2}v(h^2-v^2)\;,\\
    \langle F^{\alpha}_{8},ST_{\alpha}S^{-1}\rangle&=\frac{(N^2-4)^2}{N^4}v^3+\frac{k(2N^3-8k-N^2k-6N)}{N^4}v(h^2-v^2)\;.
    \end{align}
\end{subequations}
The force in the mixed-adjoint set is
\begin{equation}
\label{Root0mixedadjointansatzForces}
      \langle F^{\alpha}_{9},ST_{\alpha}S^{-1}\rangle=\frac{N^2-4}{N^2}v^3+\frac{k(2N^2-Nk-3)}{N^3}v(h^2-v^2)\;.
\end{equation}
Now, for a Casimir scaling, we should also equate to 0 the total force on $\psi_\alpha$ with $\alpha\cdot\beta=0$. It turns out that this is automatically satisfied after imposing the $h_{qp}$-freezing conditions  in eq. \eqref{cartanconditions}. Finally, for roots $\alpha$ such that $\alpha\cdot\beta=1$, we show the expressions for the lower order and flavor-singlet forces
\begin{subequations}
\label{Root1singletansatzForces}
\begin{align}
    \langle F^{(2)-\alpha},ST_{\alpha}S^{-1}\rangle&=h\\
    \langle F^{(3)-f-\alpha},ST_{\alpha}S^{-1}\rangle&= hv\\
    \langle F^{(3)-d-\alpha},ST_{\alpha}S^{-1}\rangle&=   \frac{N^2-4}{N^2}hv\;,\\
    \langle F^{\alpha}_{1},ST_{\alpha}S^{-1}\rangle&= (N^2-1)v^2h+2k(N-k)h(h^2-v^2)\\
    \langle F^{\alpha}_{2},ST_{\alpha}S^{-1}\rangle&=h^3\\
    \langle F^{\alpha}_{3},ST_{\alpha}S^{-1}\rangle&=\frac{h(h^2+v^2)}{2}\\
    \langle F^{\alpha}_{4},ST_{\alpha}S^{-1}\rangle&=\frac{(N-2k)^2}{N^2}h(h^2-v^2)\;,
\end{align}
\end{subequations}
 the f-adjoint forces
\begin{subequations}
\label{Root1fadjointansatzForces}
    \begin{align}
        \langle F^{\alpha}_{5},ST_{\alpha}S^{-1}\rangle&=\frac{1}{2}h(h^2+v^2)\\
    \langle F^{\alpha}_{6},ST_{\alpha}S^{-1}\rangle&=v^2h+\frac{1+Nk-k^2}{N^2}h\left(h^2-v^2\right)
    \end{align}
\end{subequations}
 the d-adjoint forces
\begin{subequations}
\label{Root1dadjointansatzForces}
    \begin{align}
        \langle F^{\alpha}_{7},ST_{\alpha}S^{-1}\rangle&=\frac{(N-2k)^2}{N^2}h(h^2-v^2)\;,\\
    \langle F^{\alpha}_{8},ST_{\alpha}S^{-1}\rangle&=\frac{N^4-8N^2+16}{N^4}v^2h+\frac{N^3k-N^2k^2-3N^2+8Nk-8k^2}{N^4}h^3\;,
    \end{align}
\end{subequations}
and the mixed-adjoint force
\begin{equation}
    \label{Root1mixedansatzForces}
    \langle F^{\alpha}_{9},ST_{\alpha}S^{-1}\rangle=\frac{Nk-k^2-1}{N^2}h(h^2-v^2)+\frac{N^2-4}{N^2}v^2h\;.
\end{equation}
Since the profile $h$ is nontrivial, there is no condition associated with the total force vanishing. However, it is important that $h$ does not depend on $k$ so as to guarantee a Casimir law. This would entail equating to $0$ the coefficients of $kh$, $k^2h$, $kh^3$, and $k^2h^3$ in the total force. Just like before, the resulting conditions are not independent of eqs. \eqref{cartanconditions}. That is, after freezing $h_{qp}$, the equation satisfied by $h$ is automatically $k$-independent  and  reads
\begin{align}
    &\nabla^2 h =
    \left(m^2+\kappa_fv+\kappa_d\frac{N^2-4}{N^2}v+\lambda_1(N^2-1)v^2+\frac{\lambda_3v^2}{2}-\lambda_4v^2+\lambda_5\frac{v^2}{2}+\lambda_6\frac{N^2-1}{N^2}v^2-\lambda_{7}v^2+\right.\nonumber\\&\left.+\lambda_{8}\frac{N^4-5N^2+16}{N^4}v^2+\lambda_{9}\frac{N^2-3}{N^2}v^2\right)h+\left(\lambda_2+\frac{\lambda_3}{2}+\lambda_4+\frac{\lambda_5}{2}+\frac{\lambda_6}{N^2}+\lambda_{7}-\frac{3}{N^2}\lambda_{8}-\frac{\lambda_{9}}{N^2}\right)h^3\;.
\end{align}
For completeness, we show also the resulting equation for the gauge field
\begin{align}
    \frac{1}{\rho}\frac{d a}{d \rho}-\frac{d^2 a}{d \rho^2}=g^2 h^2(1-a)\;.
\end{align}
These equations are those of an ANO model after an appropriate redefinition of the parameters.

\section{Stability}

  In the simple model given by eq. \eqref{higgs},  when moving from the Abelianization point at $m^2=0$, there is a neighboring region ($m^2<0$) where it becomes unstable (see Sec. \ref{Intro}). In this region, the fields prefer to align along a common direction in the Lie algebra and arbitrarily increase their norm. This way, the cubic and quartic terms are nullified and the energy due to the mass term becomes arbitrarily negative. Although we shall not analyze the parameter space in detail, we would like to note that this issue can be easily fixed in the general color and flavor symmetric setting, and even in a class of models where the 
field-content is reduced by disregarding the $\psi_q$ Higgs-sector. Moreover, this can be done while keeping the Abelian-like profiles as well as the Casimir scaling law. 

\subsection{Models with color and flavor symmetry}

For example, let us consider the model in Eqs. \eqref{gen_model},  \eqref{genV}, with $m^2$, $\kappa_f$, $\lambda_2 >0$, and $\lambda_3 > 0$ being the only nonvanishing parameters. In the new quartic contribution
\begin{align}
\lambda_2 \left(\langle \psi_1 , \psi_1\rangle^2+ \langle \psi_2 , \psi_2\rangle^2+ \dots + 2\langle \psi_1 ,\psi_2 \rangle^2 + 2\langle \psi_2 ,\psi_3 \rangle^2 + \dots \right) \;,
\label{nq}    
\end{align}
the terms with a single flavor index 
prevent the energy minimization 
with an arbitrarily large norm, thus leading to a stable model. Now, in order for the $SU(N)\to Z(N)$ SSB vacua $\psi_A=v ST_AS^{-1}$ to be preferred with respect to the trivial vacuum, the condition
\begin{align}
    m^2<\frac{2}{9}\frac{\kappa_f^2}{\lambda_2+\lambda_3}
\end{align}
must be satisfied. In addition, it can be easily seen that for 
\begin{align}
\lambda_2> \frac{\lambda_3}{N^2-2}
\end{align} 
the $SU(N)\to Z(N)$ SSB vacua are favored when compared with the aligned vacua.
In this respect, note that the mixed terms in Eq. \eqref{nq} tend to favor the orthogonality between different fields. 
Moreover, the freezing conditions for $\psi_q$ in Eqs. \eqref{firstcartancondition}-\eqref{thirdcartancondition} are satisfied at 
\[
m^2 = -\lambda_2 \left(\frac{\kappa_f}{\lambda_3}\right)^2 \;,
\]
which corresponds to $
v= -\frac{\kappa_f}{\lambda_3}$. According to the analysis in Sec. \ref{AbelianizationPoint},  this freezing automatically implies Nielsen-Olesen profiles and asymptotic Casimir scaling.

\subsection{Reduced models without $\psi_q$}
\label{Modelwitoutpsiq}

From the ensemble  point of view \cite{4densemble}, the Higgs fields $\psi_\alpha $,  $\psi_{\bar{\alpha}}$ labeled by roots are naturally associated with worldlines carrying an adjoint  charge $\alpha$. On the other hand, the  
adjoint Higgs fields $\psi_q$ labelled by Cartan indices were introduced to cope with possible matching rules in the $\mathfrak{su}(2)$ subalgebras of $\mathfrak{su}(N)$. If these matching rules were absent, it would be appropriate to limit the Higgs field-content of the effective model to $\psi_\alpha $,  $\psi_{\bar{\alpha}}$. Let us analyze what would change in  this scenario. This can be achieved  by setting $\psi_q=0$ in the energy functional and the ansatz. Of course, we do not have to worry about the conditions derived from the eqs. for $\psi_q$ (cf. \eqref{SingletCartanansatzForces}-\eqref{mixedCartanansatzForces}). In the root sector, on the other hand, eqs. \eqref{Forcerootsector1}-\eqref{Forcerootsector4} with $\psi_q=0$ are still valid and, for that reason, the ansatz still closes. The main changes are originated from eqs. \eqref{Root0singletansatzForces}-\eqref{Root0mixedadjointansatzForces} and \eqref{Root1singletansatzForces}-\eqref{Root1mixedansatzForces}, since the absence of the fields $\psi_q$ drastically modify the coefficients therein. Consequently, new conditions emerge when equating the coefficients of the new total forces on $\psi_\alpha$ to $0$ ($\alpha\cdot\beta=0$). Nevertheless, a similar analysis can be carried out and, just as before, not all conditions are independent. The freezing conditions can be chosen as
\begin{subequations}
    \begin{align}
        \kappa_f+\kappa_d + 2N^2\lambda_1v + \lambda_3v -2(\lambda_4+\lambda_7)v +  \lambda_5v+\frac{2N-3}{N}(\lambda_6+\lambda_8+\lambda_9)v=0\\
        -2N^2\lambda_1 + 4(\lambda_4+\lambda_7) -(\lambda_6+\lambda_9) -\frac{N^2+8}{N^2}\lambda_8=0  \;,
    \end{align}
\end{subequations}
while the new equation that defines $v$ is
\begin{eqnarray}
    0&=&m^2+\frac{N-2}{N}v(\kappa_f+\kappa_d)+N(N-1)v^2\lambda_1+v^2\lambda_2+\frac{N-1}{N}v^2\lambda_3+\frac{N-1}{N}v^2\lambda_5\notag\\
    &&+\frac{N^2-3N+4}{N^2}v^2\lambda_6+\frac{N^3-3N^2+4}{N^3}v^2\lambda_8+\frac{N^2-3N+2}{N^2}v^2\lambda_9  \;.
\end{eqnarray}
Again, the freezing conditions lead to a collective behavior   where the nontrivial profiles
$h_\alpha$ ($\alpha \cdot \beta =1$) are equal to a single one $h$, which satisfies a $k$-independent  Nielsen-Olesen equation
\begin{eqnarray}
    \nabla^2h &=& \left(m^2+\kappa_f\frac{N-2}{N}v+\kappa_d\frac{N-2}{N}v+\lambda_1N(N-1)v^2+ \lambda_3\frac{N-2}{2N} v^2-\lambda_4v^2+\lambda_5\frac{N-2}{2N}v^2 \right.\notag\\
    &&\left.+ \lambda_6\frac{N^2-3N+3}{N^2}v^2-\lambda_7v^2+\lambda_8\frac{N^3-3N^2+3N+4}{N^3}v^2+\lambda_9\frac{N^2-3N+3}{N^2}v^2\right)h\\
    &&+ \left(\lambda_2+ \frac{\lambda_3}{2}+\lambda_4+\frac{\lambda_5}{2} +\frac{\lambda_6}{N^2}+\lambda_7 -\frac{3\lambda_8}{N^2} -\frac{\lambda_9}{N^2}\right)h^3\;.\notag
\end{eqnarray}
In addition, when $m^2$, $\kappa_f$, $\lambda_2$, and $\lambda_3$ are the only nonvanishing parameters,  the above analysis is expected to hold for sufficiently large $\lambda_2$. In that region the favored vacua would be $\psi_\alpha = v S T_\alpha S^{-1}$, $\psi_{\bar{\alpha}} = v S T_{\bar{\alpha}} S^{-1}$. This vacuum has a lower energy than the trivial one when
\begin{align}
    m^2<\frac{2}{9 N}\kappa_f^2\frac{N^2-4N+4}{N\lambda_2+\lambda_3(N-1)}\;.
\end{align}
Moreover, 
we checked that in the region
\begin{align}
    \lambda_2>\frac{\lambda_3(N-1)}{N^3-N^2-N}\;
\end{align}
these vacua are favored with respect 
to the aligned configuration. In this example, 
the freezing condition for the fields $\psi_\alpha$ ($\alpha\cdot\beta=0$) occurs at
\begin{align}
m^2  = -\frac{\kappa_f^2}{N \lambda_3} - \frac{\lambda_2 \kappa_f^2}{\lambda_3^2} \;,
\end{align}
which corresponds to $v= -\frac{\kappa_f}{\lambda_3}$. At this point, besides stability, the reduced model displays Abelian-like vortex profiles and Casimir scaling, as the general conditions given in Sec. \ref{AbelianizationPoint} are also realized. \\

At the freezing point, in the color and flavor symmetric model and in the reduced model, the energy difference between the preferred $SU(N)\ \to Z(N)$ SSB configuration, the aligned, and the trivial one is finite. Thus, we may conclude that the $SU(N)\ \to Z(N)$ SSB pattern is stable with respect to small deviations from the freezing point.
In this case, the flux tubes only receive perturbative corrections. Also, because of the additional quartic term considered, all possible phases obtained when the mass and cubic parameters are arbitrarily varied become correctly stabilized, as the energy of the global minima will be bounded from below.

\section{Discussion}

In this work, we analyzed two classes of YMH models with a set of adjoint Higgs flavors. Initially, we considered the most general case with $SU(N)$ color and flavor symmetry constructed in terms of $N^2-1$ adjoint real scalars. Next, we also analyzed models derived from the former by disregarding Higgs flavor labels in the Cartan sector, only keeping Higgs fields labeled by the adjoint weights of $SU(N)$, which can be readily associated with the different monopole charges. In this case, the cubic and quartic interactions effectively describe the matching rules for these charges when three and four monopole worldlines meet at a point. This, together with the minimal coupling to the $SU(N)$ gauge field Goldstone modes $\Lambda_\mu$,  describe a mixed ensemble of oriented and nonoriented center vortices \cite{4densemble}. In both cases,
the $SU(N) \to Z(N)$ SSB pattern, essential to reproduce the observed $N$-ality properties of the confining states at asymptotic distances, can be realized. Here, we showed that the different properties suggested by the lattice can be accommodated in a class of models that remain stable under variations of the Higgs-field mass parameter. These properties include asymptotic Abelian profiles \cite{emt-ym}, the Casimir scaling law \cite{casimir-4d}, and the independence of the flux-tube cross-section from the $N$-ality of the quark representation \cite{stability}.
For each class, 
the generation of Abelian profiles was traced back to the possibility of freezing 
the Higgs fields having labels that are trivially transformed by Cartan transformations along the $k$-antisymmetric weights. This freezing automatically implies that
the profiles $h_\alpha$ associated to Higgs fields that do rotate under this type of transformation (there are $k(N-k)$ such fields) can be equated to a single profile $h$. The latter satisfies a Nielsen-Olesen equation that turns out to be $k$-independent. As the regularity conditions are also $k$-independent,
the above mentioned cross-section property is then implied. 
Therefore, although the models are formulated in terms of many fields, a collective behavior arises where the $k$-vortex energy is proportional to $k(N-k)$, which coincides with the quadratic Casimir of the $k$-antisymmetric representation. In both classes of models there are relatively few freezing conditions on the parameters.  In addition, for small deviations from the freezing point, the vortex properties are only perturbatively modified. Then, it is satisfying to see that properties observed or suggested in lattice simulations of $SU(N)$ YM lattice theory are ubiquitous in YMH models with adjoint flavors, which in turn provide an effective description of mixed ensembles of oriented and nonoriented center vortices also observed in the lattice.

\section*{Acknowledgments}
The Conselho Nacional de Desenvolvimento Cient\'ifico e Tecnol\'ogico (CNPq) is acknowledged for the financial support.

\appendix

\section{Weights of SU(N)}
The weights $\lambda$ of a given representation D of SU(N) are $N-1$ tuples defined in terms of the eigenvectors of the Cartan generators, as follows:
\begin{align}
    D(T_q)|\lambda\rangle =\lambda|_q |\lambda\rangle\;. 
\end{align}
When D is the fundamental(defining) representation, these weights are denoted by $\omega_i$, $i=1,\dots,N$. It is convenient to define an ordering relation for these tuples, where a given weight is said to be positive if its last nonzero component is positive. It is also convenient to define the magnetic weights $\beta_i=2N\omega_i$. Then, the magnetic weights of the defining representation are defined such that $\beta_1>\beta_2>\dots>\beta_N$. They all have the same length, i.e. $|\beta_i|^2=2(N-1)$, and different weights have the following scalar product
\begin{align}
    \beta_i\cdot\beta_j=-2\;,\;i\neq j\;.
\end{align}
Another important particular case is when D is the adjoint representation, defined by
\begin{align}
    {\rm Ad}(T_A)|_{BC}=-if_{ABC}\;.
\end{align}
The corresponding weights are known as the roots of SU(N). They are given by differences of fundamental weights, i.e., all roots can be written as
\begin{align}
    \alpha_{ij}=\omega_i-\omega_j\;,\; i\neq j\;.
\end{align}
Notice that $\alpha_{ij}$ is positive if and only if $i>j$.
\section{The structure constants of SU(N)}
\label{LieAlgebraAppendix}

This section is dedicated to recalling the definition and properties of the symmetric and antisymmetric structure constants $d_{ABC}$ and $f_{ABC}$ of SU(N). We define the antisymmetric constants in terms of the commutators
\begin{equation}
    [T_A,T_B]=if_{ABC}T_C\;.
\end{equation}
It is more elegant to define these constants in terms of an operation, which we will denote by the symbol $\wedge$, that is entirely closed in the algebra
\begin{equation}
    T_A\wedge T_B = -i[T_A,T_B]=f_{ABC}T_C\;.
\end{equation}
The actual values of the constants $f_{ABC}$ depend on a choice of basis and throughout this work we will always use the Weyl-Cartan basis which consists of $N-1$ diagonal generators $T_q\,,\, q=1,\dots,N-1$, known as the Cartan generators, and the off-diagonal generators $T_{\alpha}, T_{\bar{\alpha}}$, which are labeled by the positive roots $\alpha$ of SU(N). The off-diagonal generators are defined in terms of the root vectors $E_\alpha$, which satisfy
\begin{align}
    &[T_q,E_\alpha]=\alpha|_q E_\alpha \makebox[.5in]{,} [E_\alpha,E_{-\alpha}]=\alpha|_q T_q\;,\nonumber\\&
[E_\alpha,E_\gamma]=N_{\alpha\gamma}E_{\alpha+\gamma}\text{, for }\alpha+\gamma\neq 0\;.
\end{align}
The constant $N_{\alpha\delta}$ being zero if $\alpha+\delta$ is not a root. Then, the Hermitian off-diagonal generators are defined by
\begin{align}
    T_{\alpha}=\frac{E_\alpha+E_{-\alpha}}{\sqrt{2}} \makebox[.5in]{,}T_{\bar{\alpha}}=\frac{E_\alpha-E_{-\alpha}}{\sqrt{2}i}\;.
\end{align}
The nontrivial commutation relations in the Cartan-Weyl basis are
\begin{subequations}
\label{commutation}
\begin{align}
T_q\wedge T_{\alpha}=&\alpha|_qT_{\overline{\alpha}} \;,\\
T_q\wedge T_{\overline{\alpha}}=&-\alpha|_qT_{\alpha} \;,\\
T_{\alpha}\wedge T_{\overline{\alpha}}=&\alpha|_qT_q \;,\\
T_{\alpha}\wedge T_{\beta} =& \frac{1}{\sqrt{2}}\left( N_{\alpha,\beta}T_{\overline{\alpha+\beta}} + N_{\alpha,-\beta} T_{\overline{\alpha-\beta}} \right)\;,\\
T_{\alpha}\wedge T_{\overline{\beta}} =& \frac{1}{\sqrt{2}}\left(- N_{\alpha,\beta}T_{\alpha+\beta} + N_{\alpha,-\beta} T_{\alpha-\beta} \right)\;,\\
T_{\overline{\alpha}}\wedge T_{\overline{\beta}} =& \frac{1}{\sqrt{2}}\left(- N_{\alpha,\beta}T_{\overline{\alpha+\beta}} + N_{\alpha,-\beta} T_{\overline{\alpha-\beta}} \right)\;.
\end{align}
\end{subequations}
To evaluate the constants $f_{ABC}$, we use the identity
\begin{eqnarray}
     f_{ABC}=\langle T_A\wedge T_B,T_C\rangle\;,
\end{eqnarray}
although one caveat is worth mentioning: because of the property $T_{-\alpha}=T_{\alpha}$, $T_{\overline{-\alpha}}=-T_{\overline{\alpha}}$, the Killing products between generators associated with roots are
\begin{subequations}
\begin{align}
    \langle T_{\alpha}, T_{\beta}\rangle &= \delta_{\alpha,\beta}+\delta_{\alpha,-\beta}\;,\\
\langle T_{\overline{\alpha}},T_{\overline{\beta}}\rangle &= \delta_{\alpha,\beta}-\delta_{\alpha,-\beta}\;.
\end{align}
\end{subequations}
With this in mind, the final result for the non-zero antisymmetric constants is
\begin{subequations}
\label{fconstants}
\begin{align}
    f_{q\alpha\overline{\alpha}}&=-f_{q\overline{\alpha}\alpha}=\alpha\vert_q\;,\label{fCartan}\\
    f_{\gamma\eta\overline{\alpha}}&=\frac{1}{\sqrt{2}}\big(N_{\gamma,\eta}(\delta_{\alpha,\gamma+\eta}-\delta_{\alpha,-\gamma-\eta})+N_{\gamma,-\eta}(\delta_{\alpha,\gamma-\eta}-\delta_{\alpha,\eta-\gamma})\big)\;,\label{froots1}\\
    f_{\gamma\overline{\eta}\alpha}&=\frac{1}{\sqrt{2}}\big(-N_{\gamma,\eta}(\delta_{\alpha,\gamma+\eta}+\delta_{\alpha,-\gamma-\eta})+N_{\gamma,-\eta}(\delta_{\alpha,\gamma-\eta}+\delta_{\alpha,\eta-\gamma})\big)\;,\label{froots2}\\
    f_{\overline{\gamma}\,\overline{\eta}\,\overline{\alpha}}&=\frac{1}{\sqrt{2}}\big(-N_{\gamma,\eta}(\delta_{\alpha,\gamma+\eta}-\delta_{\alpha,-\gamma-\eta})+N_{\gamma,-\eta}(\delta_{\alpha,\gamma-\eta}-\delta_{\alpha,\eta-\gamma})\big)\;.\label{froots3}
\end{align}
\end{subequations}

In our convention, the constants $N_{\alpha,\beta}$ are given by
\begin{equation}
    |N_{\alpha,\beta}|=\begin{cases}
    \frac{1}{\sqrt{2N}},& \text{if } \vec{\alpha}+\beta\text{ is a root}\\
    0,              & \text{otherwise}
\end{cases}
\end{equation}
They also have the useful properties
\begin{subequations}
    \label{Nproperty}
    \begin{align}
        N_{-\alpha,-\beta}&=N_{\beta,\alpha}=-N_{\alpha,\beta}\;,\label{Nproperty1}\\
        N_{\alpha,\beta}&=N_{\gamma,\alpha}=N_{\beta,\gamma}\; \text{ if }\;\alpha+\beta+\gamma=0\;.\label{Nproperty2}
    \end{align}
\end{subequations}

The roots $\alpha$ and the weights $\omega$ have a few properties worth noticing:
\begin{eqnarray}
     \alpha=\omega_i-\omega_j\;,\\
     \omega_i\cdot\omega_j=\frac{N\delta_{ij}-1}{2N^2}\Rightarrow|\alpha|^2=\frac{1}{N}\;,\\
     \sum\limits_{i=1}^{N}\omega_i\vert_q\omega_i\vert_p=\sum\limits_{\alpha>0}\alpha\vert_q\alpha\vert_p=\frac{\delta_{qp}}{2}\;.
\end{eqnarray}

The symmetric constants are defined in terms of the anticommutators
\begin{equation}
    \{T_A,T_B\}=c\mathbb{I} +d_{ABC}T_C\,
\end{equation}
The appearance of a component in the direction of the identity matrix $\mathbb{I}$ comes from the fact that the anticommutator is not traceless. In fact, the constant $c$ can be found via the trace of this equation
\begin{equation}
    2\text{Tr}(T_A T_B)=c\,N\;.
\end{equation}
The basis $T_A$ is normalized in the sense of the Killing product, which can be realized as
\begin{equation}
    \langle T_A,T_B\rangle=2N\text{Tr}(T_AT_B)=\delta_{AB}\;.
\end{equation}
This leads to
\begin{equation}
    c=\frac{\langle T_A,T_B\rangle}{N^2} = \frac{\delta_{AB}}{N^2}
\end{equation}
Once again, it is more elegant to define the constants $d_{ABC}$ in terms of a product closed in the algebra. We denote this product by $\vee$ and set
\begin{equation}
    T_A\vee T_B = \{T_A,T_B\}-\frac{\langle T_A,T_B\rangle}{N^2} = d_{ABC} T_C\;. 
\end{equation}
Because the basis $T_A$ is traceless, we can also obtain these constants by
\begin{equation}
\label{dTrace}
    d_{ABC}=2N\text{Tr}(\{T_A,T_B\}T_C)\;.
\end{equation}
This expression makes clear the cyclic property $d_{ABC}=d_{BCA}$.

The constants $d_{ABC}$ have fewer interesting properties which makes it desirable to replace them with $f_{ABC}$ whenever possible. To do so, the following relations are useful \cite{Chivukula}
\begin{eqnarray}
     f_{ABE}f_{CDE}&=&d_{ACE}d_{BDE}-d_{ADE}d_{BCE}+\frac{2}{N^2}(\delta_{AC}\delta_{BD}-\delta_{AD}\delta_{BC})\;,\label{StrucConstIdent1}\\
     f_{ABE}d_{CDE}&=&d_{ADE}f_{BCE}+d_{ACE}f_{BDE}\;,\label{StrucConstIdent2}\\
     d_{AEF}d_{BEF}&=&\frac{N^2-4}{N^2}\delta_{AB}\;.\label{StrucConstIdent3}
\end{eqnarray}

Fortunately, since we are only interested in SU(N), more can be said about the symmetric constants. For that purpose, we first write the matrix realization of the Weyl-Cartan basis in terms of roots and weights
\begin{eqnarray}
     T_q\vert_{ij}&=&\omega_i\vert_{q}\delta_{ij}\;,\\
     T_{\alpha_{ab}}\vert_{ij}&=& \frac{1}{2\sqrt{N}}\left(\delta_{ia}\delta_{jb}+\delta_{ib}\delta_{ja}\right)\;,\\
      T_{\overline{\alpha}_{ab}}\vert_{ij}&=& \frac{i}{2\sqrt{N}}\left(-\delta_{ia}\delta_{jb}+\delta_{ib}\delta_{ja}\right)\;.
\end{eqnarray}
Using the components of the generators, it is possible to show 
\begin{subequations}
\label{anticommutation}
\begin{align}
T_q\vee T_p=&\sum\limits_{i=1}^N \omega_i\vert_q\omega_i\vert_p\omega_i\cdot T \;,\\
T_q\vee T_{\alpha}=&\tilde{\alpha}|_qT_{\alpha} \;,\\
T_q\vee T_{\overline{\alpha}}=&\tilde{\alpha}|_qT_{\overline{\alpha}} \;,\\
T_{\alpha}\vee T_{\overline{\alpha}}=&0 \;,\\
T_{\alpha}\vee T_{\beta} =& \frac{1}{\sqrt{2}}\left( |N_{\alpha,\beta}|T_{\alpha+\beta} + |N_{\alpha,-\beta}| T_{\overline{\alpha-\beta}} \right)\;,\\
T_{\alpha}\vee T_{\overline{\beta}} =& \frac{1}{\sqrt{2}}\left( |N_{\alpha,\beta}|T_{\overline{\alpha+\beta}} - |N_{\alpha,-\beta}| T_{\overline{\alpha-\beta}} \right)\;,\\
T_{\overline{\alpha}}\vee T_{\overline{\beta}} =& \frac{1}{\sqrt{2}}\left(- |N_{\alpha,\beta}|T_{\alpha+\beta} + |N_{\alpha,-\beta}| T_{\alpha-\beta} \right)\;.
\end{align}
\end{subequations}
For each $\alpha=\omega_i-\omega_j$, we define $\tilde{\alpha}=\omega_i+\omega_j$.

We can now use eq. \eqref{dTrace} and the analogous commutator version to evaluate the constants $d_{ABC}$ \footnote{All the roots are assumed to be positive}
\begin{subequations}
\label{dconstants}
\begin{align}
     d_{qpl}&=4N\sum\limits_{i=1}^N\omega_i\vert_q\omega_i\vert_p\omega_i\vert_l\;,\label{dCartan1}\\
    d_{q\alpha\alpha}&=d_{q\overline{\alpha}\,\overline{\alpha}}=\tilde{\alpha}\vert_q\;,\label{dCartan2}\\
    d_{\gamma\eta\alpha}&=\frac{1}{\sqrt{2}}\big(|N_{\gamma,\eta}|(\delta_{\alpha,\gamma+\eta}+\delta_{\alpha,-\gamma-\eta})+|N_{\gamma,-\eta}|(\delta_{\alpha,\gamma-\eta}+\delta_{\alpha,\eta-\gamma})\big)\;,
    \label{droots1}\\
    d_{\gamma\overline{\eta}\overline{\alpha}}&=\frac{1}{\sqrt{2}}\big(|N_{\gamma,\eta}|(\delta_{\alpha,\gamma+\eta}-\delta_{\alpha,-\gamma-\eta})-|N_{\gamma,-\eta}|(\delta_{\alpha,\gamma-\eta}-\delta_{\alpha,\eta-\gamma})\big)\;,
    \label{droots2}\\
    d_{\overline{\gamma}\,\overline{\eta}\alpha}&=\frac{1}{\sqrt{2}}\big(-|N_{\gamma,\eta}|(\delta_{\alpha,\gamma+\eta}+\delta_{\alpha,-\gamma-\eta})+|N_{\gamma,-\eta}|(\delta_{\alpha,\gamma-\eta}+\delta_{\alpha,\eta-\gamma})\big)\;.\label{droots3}
\end{align}
\end{subequations}

\end{document}